\documentclass[screen,sigconf]{acmart}
\setcopyright{cc}
\setcctype{by}
\copyrightyear{2025}
\acmYear{2025}
\acmConference[SC '25]{The International Conference for High Performance Computing, Networking, Storage and Analysis}{November 16--21, 2025}{St Louis, MO, USA}
\acmBooktitle{The International Conference for High Performance Computing, Networking, Storage and Analysis (SC '25), November 16--21, 2025, St Louis, MO, USA}
\acmDOI{10.1145/3712285.3771783}
\acmISBN{979-8-4007-1466-5/2025/11}

\usepackage[T1]{fontenc}
\usepackage{fancyhdr}
\usepackage{graphicx}
\usepackage{textcomp}
\usepackage{xcolor}
\usepackage{verbatim}
\usepackage{booktabs}
\usepackage{subcaption}
\usepackage{multirow}
\usepackage{array}
\usepackage{makecell}
\usepackage{tabularx}
\usepackage{siunitx}
\usepackage{enumitem}
\usepackage{newtxmath}
\usepackage{microtype}
\usepackage{bm}

\DeclareSIUnit\flop{FLOP}
\DeclareSIUnit\flops{FLOPs}

\usepackage{listings}

\DeclareMathAlphabet{\mathtt}{T1}{zi4}{m}{n}
\DeclareSymbolFont{incomono}{T1}{zi4}{m}{n}
\SetSymbolFont{incomono}{bold}{T1}{zi4}{b}{n}
\DeclareMathSymbol{(}{\mathopen}{incomono}{"28}
\DeclareMathSymbol{)}{\mathclose}{incomono}{"29}
\DeclareMathSymbol{+}{\mathbin}{incomono}{"2B}
\DeclareMathSymbol{-}{\mathbin}{incomono}{"2D}
\DeclareMathSymbol{=}{\mathrel}{incomono}{"3D}
\DeclareMathSymbol{,}{\mathpunct}{incomono}{"2C}
\DeclareMathSymbol{.}{\mathord}{incomono}{"2E}

\usepackage[ruled,vlined,linesnumbered]{algorithm2e}

\def\BibTeX{{\rm B\kern-.05em{\sc i\kern-.025em b}\kern-.08em
    T\kern-.1667em\lower.7ex\hbox{E}\kern-.125emX}}

\usepackage{ifpdf}
\usepackage{tikz}
\usepackage{pgfplots}
\usetikzlibrary{shapes.arrows, patterns, calc}
\usepackage{tikz-3dplot}
\usepackage{bbm}
\ifpdf
  \DeclareGraphicsExtensions{.pdf,.eps,.png,.jpg}
\else
  \DeclareGraphicsExtensions{.eps}
\fi

\usepgfplotslibrary{groupplots}
\usetikzlibrary{arrows}
\usetikzlibrary{shapes}
\usetikzlibrary{decorations.text}
\usetikzlibrary{arrows.meta}

\definecolor{RYB1}{rgb}{0.72, 0.26, 0.06}%dark rust
\definecolor{RYB2}{rgb}{0.18, 0.42, 0.41}%dark seagreen
\definecolor{RYB3}{rgb}{0.63, 0.74, 0.78}%blueish
\definecolor{RYB4}{RGB}{251,220,127}
\definecolor{RYB5}{rgb}{0.69, 0.67, 0.66}
\definecolor{RYB6}{rgb}{0.85, 0.55, 0.13}
\definecolor{RYB7}{RGB}{128, 177, 211}

\usetikzlibrary{positioning} % in the preamble
\usetikzlibrary{backgrounds,scopes}   %<------- Load libraries

\pgfdeclarelayer{background}
\pgfdeclarelayer{foreground}
\pgfsetlayers{background,main,foreground}   %% some additional layers for demo

\pgfplotsset{
    compat=1.18,
    scale only axis,
    width=0.5\textwidth,
    enlarge x limits=0.0,
    enlarge y limits=0.0,
    max space between ticks=40,
    every axis/.append style={font=\small},
    every legend/.append style={font=\small},
    every node/.append style={font=\small},  
}

\tikzset{
    every node/.append style={font=\small},
}

\definecolor{steelblue}{HTML}{A1BDC7}
\definecolor{orange}{HTML}{D98C21}
\definecolor{silver}{HTML}{B0ABA8}
\definecolor{rust}{HTML}{B8420F}
\definecolor{seagreen}{HTML}{2E6B69}
\definecolor{joshua}{HTML}{FBDC7F}
\definecolor{darksky}{HTML}{154c79}
\definecolor{purp}{RGB}{68, 14, 156}

\colorlet{lightsilver}{silver!30!white}
\colorlet{darkorange}{orange!85!black}
\colorlet{darksilver}{silver!85!black}
\colorlet{darksteelblue}{steelblue!85!black}
\colorlet{darkrust}{rust!85!black}
\colorlet{darkseagreen}{seagreen!85!black}

\pgfplotscreateplotcyclelist{newcolors}{
{darksky,every mark/.append style={fill=darksky,mark size={2.5}},mark=*},
{seagreen,every mark/.append style={fill=seagreen},mark=square*},
{steelblue,every mark/.append style={fill=steelblue,mark size={3}},mark=triangle*},
{silver,every mark/.append style={fill=silver,mark size={3}},mark=diamond*},
{joshua,every mark/.append style={fill=joshua,mark size={3}},mark=pentagon*},
{orange,every mark/.append style={fill=orange,mark size={4}},mark=10-pointed star},
{rust,every mark/.append style={fill=rust},mark=*},
}

\newcommand{\btau}{\boldsymbol{\tau}}
\newcommand{\bu}{\mathbf{u}}
\newcommand{\dd}{\mathrm{d}}

\hypersetup{
    colorlinks=true,
    linkcolor=darkrust,
    citecolor=darkseagreen,
    urlcolor=darksilver
}

\usepackage{cleveref}
\crefname{lstlisting}{listing}{listings}
\Crefname{lstlisting}{Listing}{Listings}

\usepackage{mathtools}
\usetikzlibrary{external}
\SetCommentSty{mycommfont}

\newcommand\blfootnote[1]{%
  \begingroup
  \renewcommand\thefootnote{}\footnote{#1}%
  \addtocounter{footnote}{-1}%
  \endgroup
}

\begin{document}

\citestyle{acmnumeric}

\title{Simulating many-engine spacecraft: Exceeding 1~quadrillion \\ degrees of freedom via information~geometric regularization}

\author{Benjamin Wilfong}
\affiliation{%
  \institution{Georgia Institute of Technology}
  \city{Atlanta}
  \state{Georgia}
  \country{USA}
}

\author{Anand Radhakrishnan}
\affiliation{%
  \institution{Georgia Institute of Technology}
  \city{Atlanta}
  \state{Georgia}
  \country{USA}
}

\author{Henry Le Berre}
\affiliation{%
  \institution{Georgia Institute of Technology}
  \city{Atlanta}
  \state{Georgia}
  \country{USA}
}

\author{Daniel J. Vickers}
\affiliation{%
  \institution{Georgia Institute of Technology}
  \city{Atlanta}
  \state{Georgia}
  \country{USA}
}

\author{Tanush Prathi}
\affiliation{%
  \institution{Georgia Institute of Technology}
  \city{Atlanta}
  \state{Georgia}
  \country{USA}
}

\author{Nikolaos Tselepidis}
\affiliation{%
  \institution{NVIDIA}
  \city{Zurich}
  \country{Switzerland}
}

\author{Benedikt Dorschner}
\affiliation{%
  \institution{NVIDIA}
  \city{Zurich}
  \country{Switzerland}
}

\author{Reuben Budiardja}
\affiliation{%
  \institution{Oak Ridge National Lab}
  \city{Oak Ridge}
  \state{Tennessee}
  \country{USA}
}
\authornote{This manuscript has been authored in part by UT-Battelle, LLC, under contract DE-AC05-00OR22725 with the US Department of Energy (DOE). The US government retains and the publisher, by accepting the article for publication, acknowledges that the US government retains a nonexclusive, paid-up, irrevocable, worldwide license to publish or reproduce the published form of this manuscript, or allow others to do so, for US government purposes. DOE will provide public access to these results of federally sponsored research in accordance with the \href{http://energy.gov/downloads/doe-public-access-plan}{DOE Public Access Plan}.}

\author{Brian Cornille}
\affiliation{%
  \institution{Advanced Micro Devices}
  \city{Naperville}
  \state{Illinois}
  \country{USA}
}

\author{Stephen Abbott}
\affiliation{%
  \institution{Hewlett Packard Enterprise}
  \city{St. Paul}
  \state{Minnesota}
  \country{USA}
}

\author{Florian Sch\"afer}
\affiliation{%
  \institution{New York University}
  \city{New York}
  \state{New York}
  \country{USA}
}
\authornote{Equal contribution}
\email{florian.schaefer@nyu.edu}

\author{Spencer H. Bryngelson}
\affiliation{%
  \institution{Georgia Institute of Technology}
  \city{Atlanta}
  \state{Georgia}
  \country{USA}
}
\authornotemark[2]
\email{shb@gatech.edu}

\renewcommand{\shortauthors}{Wilfong et al.}

\begin{abstract}
We present an optimized implementation of the recently proposed information geometric regularization (IGR) for unprecedented scale simulation of compressible fluid flows applied to multi-engine spacecraft boosters.
We improve upon state-of-the-art computational fluid dynamics (CFD) techniques in terms of computational cost, memory footprint, and energy-to-solution metrics.
Unified memory on coupled CPU--GPU or APU platforms increases problem size with negligible overhead. 
Mixed half/single-precision storage and computation are used on well-conditioned numerics.
We simulate flow at 200~trillion grid points and 1~quadrillion degrees of freedom, exceeding the current record by a factor of 20.
A factor of 4 wall-time speedup is achieved over optimized baselines.
Ideal weak scaling is observed on OLCF Frontier, LLNL El~Capitan, and CSCS Alps using the full systems. 
Strong scaling is near ideal at extreme conditions, including 80\% efficiency on CSCS~Alps with an 8~node baseline and stretching to the full system.
\end{abstract}




\maketitle

\section{Justification for ACM Gordon Bell Prize}

Largest compressible computational fluid dynamics simulation, exceeding 1~quadrillion degrees of freedom and a factor of 20 beyond state-of-the-art.
Enabled by a large-scale optimized use of inviscid (information geometric) regularization.
Time-to-solution and energy-to-solution decrease by up to a factor of 4 and 5.4 in double precision, respectively, with greater decreases in single and mixed precision.
\blfootnote{Code available at \url{https://github.com/MFlowCode/MFC}}

\section{Performance Attributes}

\begin{table}[ht]
    \centering
    \caption{Summary of Performance Attributes}
    \label{t:perf}
    \small
    \begin{tabularx}{\columnwidth}{@{}lX@{}}
        \toprule
        Performance attribute & This submission \\ 
        \midrule
        Category of Achievement & Scalability, problem size, time-to-solution \\
        Type of Method Used & Finite volume w/ information regularization \\
        Results Reported Based On & Whole application including I/O \\
        Precision Reported & Mixed (FP16/32), Single, Double \\
        System Scale & Full system \\
        Measurement Mechanism & Timers, FLOPs, Power management counters \\
        \bottomrule
    \end{tabularx}
\end{table}

\section{Overview of the Problem}

The 21st century witnesses a ``new space race''~\cite{pekkanen2019governing} driven by private companies replacing government agencies in providing launch services. 
The resulting drop in launch cost enabled many business models, ranging from satellites to space manufacturing~\cite{denis2020new}.
The new emphasis on cost efficiency motivated innovations in rocket design, such as reusing rocket stages and leveraging economies of scale.

An example of such innovations is the use of many-engine rockets. 
Five large F-1 engines powered the first stage of the Saturn~V rocket. 
Their size was not constrained by road width or transportation logistics but dictated by engineering requirements.
Instead, SpaceX Super~Heavy, the first stage of Starship, is powered by 33 smaller Raptor engines. 
This has multiple advantages. 
The economy of scale benefits the production of a larger number of smaller engines, and their relatively compact size allows their transport via standard road infrastructure.
The multitude of engines also provides a degree of redundancy, and a small number of engine failures can be compensated for without risking mission success.
Upon reuse, the defective engines can be replaced.
Further, when landing the Super~Heavy after flight, thrust can be reduced by turning off individual engines.

Large arrays of small engines create new design challenges.
The exhaust plumes of densely packed engines can interact, propelling hot gas toward the rocket base and heating it.
This so-called base heating can cause mission failure~\cite{mehta2013numerical}, mandating heat shields on the rocket base that increase weight and cost.

Mitigating base heating most cost-effectively, in terms of both dollars and weight, requires understanding the mechanism by which engine exhaust is reflected towards the rocket and identifying which parts are most affected.
A key difficulty in experimental approaches to understanding plume recirculation is that it depends on numerous parameters, including varying ambient pressure as the rocket traverses the atmosphere and engine thrust vectoring for steering.
A detailed flow field characterization under a broad range of conditions is only feasible with numerical simulations. 
They even allow probing the impacts of changes to the rocket design.
Prior work on the numerical simulation of interacting rocket plumes was limited to small numbers (up to 7) of rocket engines and limited resolution (up to $\approx{}10$ million grid points)~\cite{mehta2013numerical,wang2024effect,ren2024numerical}. 
Our work addresses this shortcoming by utilizing the latest flagship exascale systems, leveraging their hardware design and coupling novel algorithms, computational methods, and the optimizations they enable.
With this, we can simulate the interaction of rocket engine plumes at unprecedented scales.

As a model, we represent the exhaust via the compressible Navier--Stokes equations
\begin{align}
    \label{eqn:mabalance}
    \frac{\partial \rho}{\partial t} + \nabla \cdot (\rho \bu) &= 0, \\
    \label{eqn:mobalance}
    \frac{\partial (\rho \bu)}{\partial t} + \nabla \cdot (\rho \bu \otimes \bu + p \mathbf{I} - \btau) &= \mathbf{0}, \\
    \label{eqn:enbalance}
    \frac{\partial E}{\partial t} + \nabla \cdot \left[(E + p)\bu - \bu \cdot \btau \right] &= 0,
\end{align}
with the ideal gas law equation of state
\begin{equation}
    \label{eqn:state_eqn}
    p = (\gamma - 1)\rho e, 
    \quad \text{for} \quad 
    e \coloneqq E/\rho - \|\bu\|^2 / 2
\end{equation}
and, for $\mu, \zeta$ denoting shear, bulk viscosity, constitutive law 
\begin{equation}
    \label{eqn:constitutive_law}
    \tau_{ij} = \mu\left(\frac{\partial u_i}{\partial x_j} + \frac{\partial u_j}{\partial x_i}\right) + \left(\zeta - {2}\mu/{3}\right) \delta_{ij} \frac{\partial u_i}{\partial x_j}.
\end{equation}
\Cref{eqn:mabalance,eqn:mobalance,eqn:enbalance} describe the conservation of mass, momentum, and energy, and $\btau$ is the viscous stress tensor and $p$ the pressure.
Tracking the mixture ratios of different gases and fluids, as well as their chemical reactions, is a natural extension of the demonstration.
For this work and its implications, we focus on \cref{eqn:mabalance,eqn:mobalance,eqn:enbalance,eqn:state_eqn,eqn:constitutive_law}.

\subsection*{Summary of Contributions} 
\begin{itemize}[leftmargin=*]
    \item Information geometric regularization foregoes nonlinear viscous shock capturing, enabling linear off-the-shelf numerical schemes and sequential summation of right-hand side contributions.
    \item Unified addressing on tightly coupled CPU--GPU and APU platforms increases total problem size with negligible performance hit. 
    \item FP32 compute and FP16 storage further reduce memory use while remaining numerically stable, enabled by the algorithm's well-conditioned numerics.
    \item Reduce memory footprint 25-fold over state-of-the-art.
    \item Improve time and energy-to-solution factors of 4 and 5.4, compared to an optimized implementation of state-of-the-art methods.
    \item First CFD simulation exceeding 200T grid points and 1~quadrillion degrees of freedom, improving on previous largest simulations by a factor of 20.
\end{itemize} 

\subsection*{Example simulation result}

\begin{figure}
    \centering
    \includegraphics[width=\columnwidth]{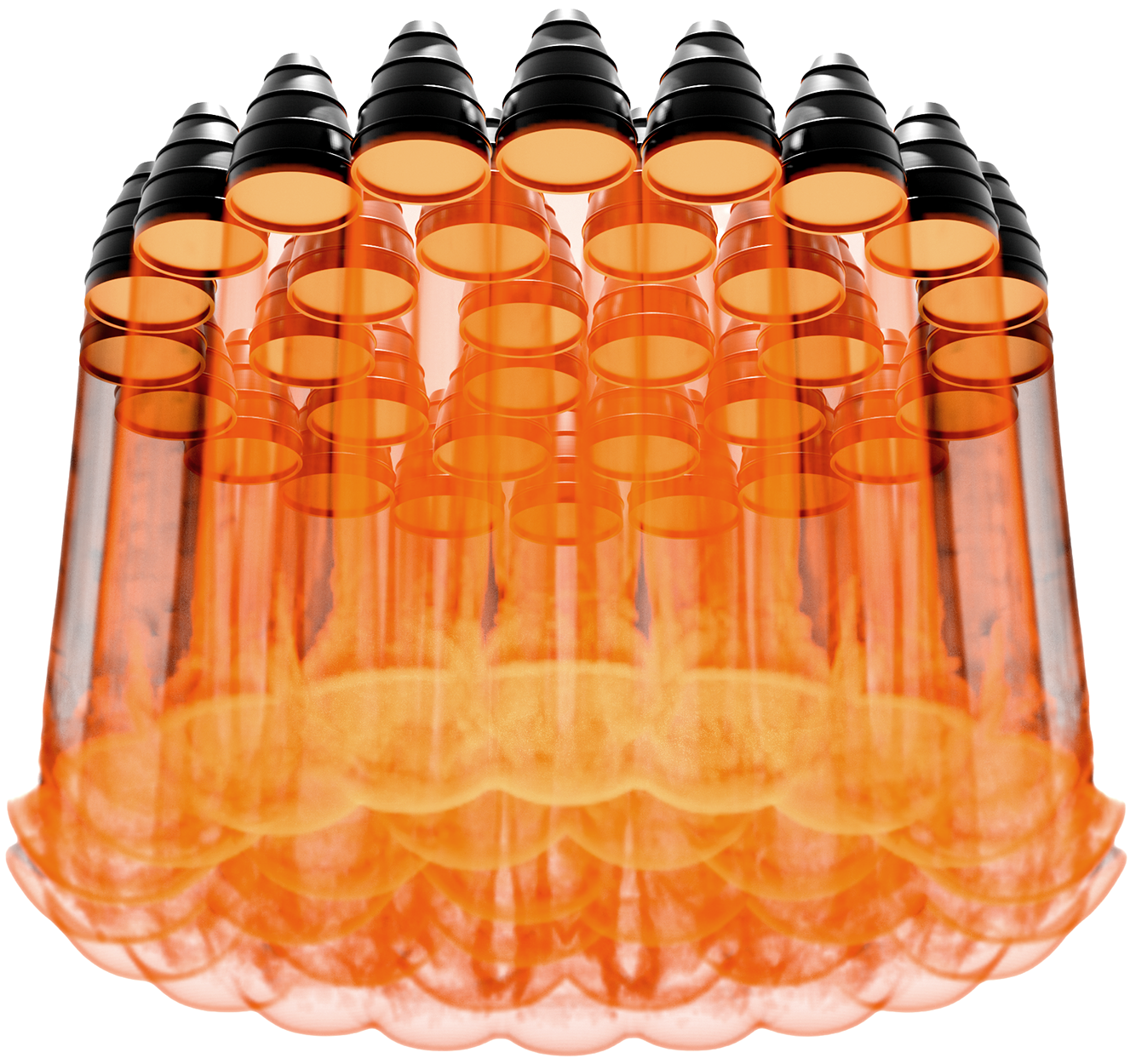}
    \Description{A rendering of 33 interacting rocket thrusters.}
    \caption{
    Simulation results showing the interacting plumes from an array of 33 thrusters in a configuration inspired by the SpaceX Super Heavy with 16.5 trillion degrees of freedom. 
    (The thrusters themselves are for visualization purposes.
    We model them through inflow boundary conditions.)
    }
    \label{fig:example}
\end{figure}

\Cref{fig:example} shows a visualization for a simulation run on CSCS Alps.
The visualization shows the interactions between an array of 33 Mach~10 rocket engines organized in an array inspired by that of the SpaceX Super Heavy, each resolved by 600~grid~cells across its outlet.
The high resolution allows fine-scale details to be represented despite the high Mach number.
The simulation of \cref{fig:example} uses a rectilinear grid of 3.3T cells and ran for 16~hours on 9.2K~GH200s (2300~Alps nodes).

\section{Current State of the Art}

\subsection{Shock capturing}
\label{sec:sota_shock_capturing}

\subsubsection*{Shock waves}

Shock waves are the primary concern in high-speed compressible fluid dynamics simulations.
They arise in numerous natural and man-made phenomena, including supernovae, air, and spacecraft.
The velocity and density fields of compressible high-speed flows sharpen over time, eventually forming macroscopic discontinuities or shock waves.
However, on the microscopic scale, the gas viscosity balances the steepening of shock waves, resulting in smooth profiles. 
In practice, this happens on the scale of the mean free path of gas particles, orders of magnitude smaller than the quantities of interest.

\subsubsection*{Computation with shocks: A multiscale problem}

Higher-order numerical methods exploit the target solution's regularity to approximate it with smooth functions, most frequently polynomials.
The mean-free path is many orders of magnitude shorter than a realistic computational grid spacing. 
Thus, shocks appear as discontinuities on the grid scale, and the direct application of higher-order methods leads to Gibbs--Runge oscillations and subsequent simulation failure.
\emph{Shock capturing} modifies either the equation or its discretization to obtain a well-behaved object on the grid scale.  
It amounts to coarse-graining the microscopic shock and correctly representing its macroscopic effects without resolution at the grid level.
The resulting coarse-scale model should preserve smooth grid-scale oscillations due to turbulence, reactions, or acoustic effects.

\subsubsection*{Existing approaches}

Artificial viscosity mitigates Gibbs--Runge oscillations at the cost of excessive dissipation of fine-scale features. 
To remedy the latter, numerous approaches apply artificial diffusivity \emph{adaptively}, in the vicinity of the shock~\cite{cook2004high,guermond2011entropy,mani2009suitability}.

Realizing a viscous regularization is challenging in practice.
A lack of viscosity creates spurious oscillations, while excessive viscosity dissipates the solution.
Common choices, such as the localized artificial diffusivity (LAD) of~\cite{cook2004high}, attempt to strike this balance.
Viscous methods spread the shock over multiple grid points, but the resulting shock profile is not high-order smooth (\cref{fig:igr_lad}~(a,i)). 
This can still lead to Gibbs--Runge oscillations overcoming the regularization and destabilizing solutions.
Increasing the shock width requires increasing the strength of the artificial viscosity, which smears true physical features critical to representing the flow (\cref{fig:igr_lad}, (b,i)).
In the presence of sufficiently strong shocks, the required artificial viscosity affects the CFL numbers of the explicit time steppers considered state-of-the-art for such hyperbolic problems.

Limiters are an alternative to artificial viscosity. 
They adaptively lower the order of the numerical method at shocks~\cite{van1979towards,liu1994weighted}.
They are more robust but also risk dissipating fine-scale features.
Riemann solvers aim to mitigate this problem but add computational cost~\cite{toro2019hllc,roe1981approximate,harten1983upstream}.

\subsection{GPU memory}

The evolution of GPU memory over the past decades, in terms of capacity and bandwidth, has been highly beneficial for CFD applications, which exhibit low arithmetic intensity.
Thus, performance is limited by memory bandwidth.
The spatio-temporal scale separation in fluid flows requires a high resolution, necessitating large amounts of memory to be accessed at high speeds.
Both NVIDIA and AMD have blurred the lines between GPU and CPU memory by introducing coherent CPU-to-GPU interfaces: \SI{900}{\giga\byte\per\second} bidirectional bandwidth for the NVIDIA Grace~Hopper and InfinityFabric at $8\times 72\,\si{\giga\byte\per\second}$ for the AMD Trento+4~MI250X configuration of Frontier.
These fast interconnects, along with technologies such as unified virtual memory (UVM), allow CPU memory for GPU computations and even (in the case of Grace~Hopper) allow the GPU to saturate the host memory bandwidth.
Furthermore, AMD introduced the MI300A in LLNL's El~Capitan and Tuolumne, with a single physical HBM pool accessed by both CPU and GPU devices.
This strategy eliminates concerns about device and host pointers or memory storage locations.

\subsection{Floating point computation}\label{sec:mixed}

The advent of artificial intelligence has fueled the development of algorithms with reduced and mixed precision, although their adoption in HPC applications remains limited.
Most traditional scientific applications rely on FP64 computation and storage.
However, recent advances in software and numerical methods demonstrate the viability of sub-FP64 precision for well-conditioned numerical methods and a range of traditional fluid flow problems~\cite{karp2024sensitivity}.
Still, FP64 is the de facto standard for compressible, shock-laden flow simulation and traditional shock-capturing techniques.
WENO-based reconstruction methods and approximate Riemann solves, considered state-of-the-art and used as the baseline in this work, involve poorly conditioned operations and thus are not well suited to reduced precision~\cite{Brogi2024OpenFOAMPrecision}.
We elide such issues herein by avoiding numerical shock capturing entirely via IGR.

\subsection{Large CFD simulations}

Previous Gordon~Bell winner \cite{rossinelli201311} performed a 10T grid point CFD simulation on IBM~BlueGene/Q.
The system's substantial CPU memory enabled this simulation.
However, the wall time required to process such a large simulation makes achievable physical time scales short.
More recently, \cite{sathyanarayana2025high} solved a compressible CFD problem of 10T grid points when extrapolating to 100\% of Frontier.

The current largest accelerator-based flow simulations are for incompressible flows, which have fewer degrees of freedom.
The largest simulation represented turbulence on a 30T point grid using OLCF~Frontier~\cite{yeung2025gpu}.
Large time step costs arise from all-to-all communications, which dominate the time-to-solution.

\section{Innovations Realized}

\subsection{The need for scale}

Current CFD simulations struggle to resolve phenomena across strongly separated space and time scales.
The fluid dynamics of engineering interest involve these multiscale interactions; in the most elementary case, large coherent flow structures and small turbulent eddies.

For external aerodynamics, such as the jets we focus on in this study, high-fidelity predictions require faithful representation of the interaction between large-scale wake structures and small vortices.
Current methods force compromises.
Fine grids are required to represent shock waves, acoustic phenomena, and their turbulent interactions.
Current state-of-the-art methods suffer from numerical dissipation, dissipating important flow features.
By combining high resolution with low wall~time cost and an \emph{inviscid} regularization (IGR), we reduce cell sizes relative to flow feature scales, overcoming these challenges.

\begin{figure}
    \centering
    \tikzsetnextfilename{igr_lad}
    \begin{tikzpicture}
    \begin{groupplot}[
	group style={group size=2 by 2,
	horizontal sep=0.55cm,
        vertical sep=0.5cm,},
    ]
    \nextgroupplot[
        ylabel={$p(x)$},
        height=0.14\textwidth,
        width=0.20\textwidth,
        ymin=0.0,
        ymax=1.0,
        xmin=0.875,
        xmax=0.95,
        yticklabels={{$0$}},
        xmajorticks=false,
        xtick={0.87, 0.9, 0.93},
        ymajorticks=false,
        enlarge x limits=0.,
        enlarge y limits=0.,
        legend style={
            at={(1.015,1.0)},inner sep=3pt,anchor=south,legend columns=3,
            legend cell align={left}, draw=none,fill=none},
	]	
    
        \addlegendimage{steelblue,ultra thick}
        \addlegendimage{orange,ultra thick}
        \addlegendimage{rust,ultra thick}
    
        \legend{Exact, LAD (current SoA), IGR (this work)}

        \addplot[steelblue,ultra thick] 
            table[x index={0},y index={1}, col sep=space, each nth point=5, 
            filter discard warning=false, unbounded coords=discard] 
            {data/igr_lad/ref_shock_data.csv};
        \addplot[orange,ultra thick] 
            table[x index={0},y index={3}, col sep=space, each nth point=5, 
            filter discard warning=false, unbounded coords=discard]
            {data/igr_lad/lad_shock_data.csv};
        
    \nextgroupplot[
        ylabel={},
        height=0.14\textwidth,
        width=0.20\textwidth,
        ymin=0.3725,
        ymax=0.4225,
        xmin=0.0,
        xmax=1.0,
        yticklabels={{$0$}},
        xmajorticks=false,
        xtick={0.2, 0.5, 0.8},
        ymajorticks=false,
        enlarge x limits=0.,
        enlarge y limits=0.,
	]	
        \addplot[steelblue,ultra thick] 
            table[x index={0},y index={1}, col sep=space, each nth point=5, 
            filter discard warning=false, unbounded coords=discard]
            {data/igr_lad/ref_osc_data.csv};
        \addplot[orange,ultra thick] 
            table[x index={0},y index={3}, col sep=space, each nth point=5, 
            filter discard warning=false, unbounded coords=discard] 
            {data/igr_lad/lad_osc_data.csv};

    \nextgroupplot[
        xlabel={$x$},
	ylabel={$p(x)$},
	height=0.14\textwidth,
	width=0.20\textwidth,
        ymax=1.0,
        xmin=0.875,
        xmax=0.95,
        yticklabels={{$0$}},
        xmajorticks=false,
        xtick={0.87, 0.9, 0.93},
        ymajorticks=false,
        yminorticks=false,
        ymin=0,
	]	
        \addplot[steelblue,ultra thick] 
            table[x index={0},y index={1}, col sep=space, each nth point=5, 
            filter discard warning=false, unbounded coords=discard]
            {data/igr_lad/ref_shock_data.csv};
        \addplot[rust,ultra thick] 
            table[x index={0},y index={3}, col sep=space, each nth point=5,
            filter discard warning=false, unbounded coords=discard]
            {data/igr_lad/igr_shock_data.csv};
        
    \nextgroupplot[
        xlabel={$x$},
        height=0.14\textwidth,
        width=0.20\textwidth,
        ymin=0.3725,
        ymax=0.4225,
        xmin=0.0,
        xmax=1.0,
        yticklabels={{$0$}},
        xmajorticks=false,
        xtick={0.2, 0.5, 0.8},
        ymajorticks=false,
	]	
        \addplot[steelblue,ultra thick] 
            table[x index={0},y index={1}, col sep=space, each nth point=5, 
            filter discard warning=false, unbounded coords=discard]
            {data/igr_lad/ref_osc_data.csv};
        \addplot[rust,ultra thick]
            table[x index={0},y index={3}, col sep=space, each nth point=5,
            filter discard warning=false, unbounded coords=discard]
            {data/igr_lad/igr_osc_data.csv};
            
\end{groupplot}

\node[anchor=north, yshift=-2ex] at (group c1r2.south) {(a) Shock problem};
\node[anchor=north, yshift=-2ex] at (group c2r2.south) {(b) Oscillatory problem};
\node[rotate=90] at ($ (group c1r1.west) $) [yshift=0.8cm]
  {(i) LAD v.\ Exact};
\node[rotate=90] at ($ (group c1r2.west) $) [yshift=0.8cm]
  {(ii) IGR v.\ Exact};

\end{tikzpicture}
    \Description{Plots showing a comparison between the solutions given by local artificial diffusion methods and IGR for discontinuous and oscillatory solutions.}
    \caption{
        Inviscid regularization: Localized artificial diffusion (LAD) spreads shocks over a user-defined width (a,i).
        The resulting curve is not high-order smooth. 
        This can cause methods with a high order of accuracy to develop oscillations and, ultimately, fail.
        Increasing the width for coarser discretizations yields unphysical and significant dissipation of oscillatory solution profiles (b,i).
        Information geometric regularization replaces shocks with smooth profiles (a,ii) at the grid scale and preserves oscillatory features (b,ii).
    }
    \label{fig:igr_lad}
\end{figure}

\subsection{Shock treatment with information geometric regularization}

We avoid the limitations of standard shock-capturing approaches via the first inviscid regularization of the equations, called \emph{information geometric regularization} (IGR) as recently proposed by~Cao and Sch\"afer~\cite{cao2023information}.
IGR was first derived in the pressureless (infinite Mach number) case, where shocks amount to the loss of injectivity of the flow map $x \mapsto \phi_t(x)$ that maps gas particles from their initial position to their position at time $t$. 
IGR modifies the geometry according to which the flow map evolves, such that particle trajectories $t \mapsto \phi_t(x_0)$ do not cross and instead asymptotically approach each other (see \cref{fig:regularization}). 
This preserves the long-time post-shock behavior, prevents the formation of grid-level singularities, and recovers the nominal vanishing viscosity solutions in the limit.
This result has been proven rigorously in the unidimensional, pressureless case~\cite{cao2024information}.

Applying IGR to the compressible Euler equations amounts to adding a so-called \emph{entropic pressure} $\Sigma$ to $p$, resulting in the modified conservation law:
\begin{align}
    \label{eqn:igrmabalance}
    \frac{\partial \rho}{\partial t} + \nabla \cdot (\rho \bu) &= 0,\\
    \label{eqn:igrmobalance}
    \frac{\partial (\rho \bu)}{\partial t} + \nabla \cdot (\rho \bu \otimes \bu + (p + \Sigma) \mathbf{I} - \btau) &= \mathbf{0},\\
    \label{eqn:igrenbalance}
    \frac{\partial E}{\partial t} + \nabla \cdot \left[(E + p + \Sigma)\bu - \bu \cdot \btau \right] &= 0,\\
    \label{eqn:igrelliptic}
    \alpha\left[\operatorname{tr}{(\nabla \bu)^2} + \operatorname{tr}^2{(\nabla \bu)} \right]&= \frac{\Sigma}{\rho} - \alpha \nabla \! \cdot \! \left(\frac{\nabla \Sigma}{\rho}\right)\!.
\end{align}
Shown in \cref{fig:igr_lad}, this \emph{inviscid} regularization yields smooth solutions without diffusion of fine-scale features.
The parameter $\alpha \propto \Delta x^2$, where $\Delta x$ is the nominal mesh spacing, determines the width of the smoothly expanded shocks.
Computing the flux requires solving the auxiliary equation \cref{eqn:igrelliptic}, but $\sqrt{\alpha}$ is proportional to the mesh size.
So, the resulting discrete system is a uniformly well-conditioned grid-point-local problem.
Using the previous solution as a warm start, $\lessapprox 5$ Jacobi or Gauss--Seidel sweeps per flux computation suffice at negligible computational cost. 

\begin{figure}
    \centering
    \tikzsetnextfilename{regularization}
    \begin{tikzpicture}

\begin{groupplot}[
    compat=1.3,
    group style={group size=2 by 1,
    horizontal sep=0.55cm},
]
\nextgroupplot[
    xlabel={$\phi_(x)$},
    ylabel={$t$},
    height=0.2025\textwidth,
    width=0.2025\textwidth,
    ymin=-0.05,
    ymax=1.0,
    xmin=-0.05,
    xmax=1.0,
    xticklabels={{$x_1$}, {$x_2$}},
    yticklabels={{$0$}},
    xtick={0,0.5},
    ytick={0},
    legend style={
        at={(1.10,1.00)},inner sep=3pt,anchor=south,
        legend columns=5,legend cell align={left},
        draw=none,fill=none,
        /tikz/every even column/.append style={column sep=0.12cm}},
    ]	

    \addlegendimage{empty legend}
    \addlegendentry{$\alpha = $}
    \addlegendimage{steelblue,very thick}
    \addlegendimage{joshua,very thick}
    \addlegendimage{orange,very thick}
    \addlegendimage{rust,very thick}

    \addlegendentry{$0.0$ (Exact)};
    \addlegendentry{$10^{-5}$};
    \addlegendentry{$10^{-4}$};
    \addlegendentry{$10^{-3}$};
    
    %         $\alpha = 10^{-5}$, $\alpha = 10^{-4}$, $\alpha = 10^{-3}$
    % \legend{
    %     }

    \addplot[rust,very thick] 
        table[x index={1},y index={0}, col sep=comma] 
        {data/igr_trajectories/characteristics_alpha_0.001.txt};

    \addplot[rust,very thick] 
        table[x index={2},y index={0}, col sep=comma, each nth point=100, 
        filter discard warning=false, unbounded coords=discard] 
        {data/igr_trajectories/characteristics_alpha_0.001.txt};
    \addplot[orange,very thick] 
        table[x index={1},y index={0}, col sep=comma, each nth point=100,
        filter discard warning=false, unbounded coords=discard] 
        {data/igr_trajectories/characteristics_alpha_0.0001.txt};

    \addplot[orange,very thick] 
        table[x index={2},y index={0}, col sep=comma] 
        {data/igr_trajectories/characteristics_alpha_0.0001.txt};

    \addplot[joshua,very thick] 
        table[x index={1},y index={0}, col sep=comma, each nth point=100, 
        filter discard warning=false, unbounded coords=discard]
        {data/igr_trajectories/characteristics_alpha_1.0e-5.txt};

    \addplot[joshua,very thick] 
        table[x index={2},y index={0}, col sep=comma, each nth point=100, 
        filter discard warning=false, unbounded coords=discard]
        {data/igr_trajectories/characteristics_alpha_1.0e-5.txt};
    \addplot[steelblue,very thick] 
        table[x index={1},y index={0}, col sep=comma, each nth point=100, 
        filter discard warning=false, unbounded coords=discard ]
        {data/igr_trajectories/characteristics_alpha_0.0.txt};
    \addplot[steelblue,very thick] 
        table[x index={2},y index={0}, col sep=comma, 
        each nth point=100, filter discard warning=false, unbounded coords=discard ] {data/igr_trajectories/characteristics_alpha_0.0.txt};
    
    \draw[thick, darksilver] (axis cs: 0.575,0.55) rectangle (axis cs: 0.675,0.45);

    \draw[darksilver] (axis cs: 0.0,0.0) -- (axis cs: 1.5,0.0);

    \coordinate (tr1) at (axis cs:0.675,0.55);
    \coordinate (br1) at (axis cs:0.675,00.45);

\nextgroupplot[
    xlabel={$\phi_t(x)$},
    height=0.2025\textwidth,
    width=0.2025\textwidth,
    ymin=0.45,
    ymax=0.55,
    xmin=0.575,
    xmax=0.675,
    xticklabels={{$\phantom{x_1}$}},
    yticklabels={},
    xtick={0.575},
    ymajorticks=false,
    enlarge x limits=0.,
    enlarge y limits=0.,
    ]	

    \addplot[rust,very thick] table[x index={1},y index={0}, col sep=comma] 
        {data/igr_trajectories/characteristics_alpha_0.001.txt};

    \addplot[rust,very thick] table[x index={2},y index={0}, col sep=comma] 
        {data/igr_trajectories/characteristics_alpha_0.001.txt};

    \addplot[orange,very thick] table[x index={1},y index={0}, col sep=comma] 
        {data/igr_trajectories/characteristics_alpha_0.0001.txt};

    \addplot[orange,very thick] table[x index={2},y index={0}, col sep=comma] 
        {data/igr_trajectories/characteristics_alpha_0.0001.txt};

    \addplot[joshua,very thick] table[x index={1},y index={0}, col sep=comma] 
        {data/igr_trajectories/characteristics_alpha_1.0e-5.txt};

    \addplot[joshua,very thick] table[x index={2},y index={0}, col sep=comma] 
        {data/igr_trajectories/characteristics_alpha_1.0e-5.txt};

    \addplot[steelblue,very thick] table[x index={1},y index={0}, col sep=comma] 
        {data/igr_trajectories/characteristics_alpha_0.0.txt};

    \addplot[steelblue,very thick] table[x index={2},y index={0}, col sep=comma] 
        {data/igr_trajectories/characteristics_alpha_0.0.txt};

    \coordinate (tl1) at 
            (axis cs:\pgfkeysvalueof{/pgfplots/xmin},\pgfkeysvalueof{/pgfplots/ymax});
    \coordinate (bl1) at 
            (axis cs:\pgfkeysvalueof{/pgfplots/xmin},\pgfkeysvalueof{/pgfplots/ymin});

    \end{groupplot}

    \draw[thick, darksilver] (tr1) -- (tl1);
    \draw[thick, darksilver] (br1) -- (bl1);
\end{tikzpicture}
    \Description{Plots showing the modified geometry of shocks when applying IGR.}
    \caption{
        Information geometric regularization modifies shocks by changing the geometry according to which the flow map $\phi_t$ evolves in time.
        In the modified geometry, the trajectories of two tracer particles $t \mapsto \phi_t(x_1), \phi_t(x_2)$ converge in $t$ rather than cross.
        The regularization strength $\alpha$ determines the rate of convergence.
        The vanishing viscosity solution is recovered in the $\alpha \rightarrow 0$ limit.
        Figure adapted from Cao and Sch\"afer~\cite{cao2023information} with author permission.
    }
    \label{fig:regularization}
\end{figure}

\subsubsection*{Discretization}

IGR allows one to bypass shock capturing, instead using a third- or fifth-order accurate finite volume method directly.
Lax--Friedrichs numerical fluxes treat the hyperbolic part of the equation. 
Due to the high Reynolds number of the problem under consideration, we find that a second-order accurate approximation of the derivatives of $\bu$ suffices to compute the viscous stress tensor. 
We reuse these derivatives to compute the left side of \cref{eqn:igrelliptic} and discretize the elliptic operator on the right using a standard 7-point stencil. 

For each computation of the hyperbolic flux, we solve \cref{eqn:igrelliptic} using up to 5~sweeps of Jacobi or Gauss--Siedel iteration, with the previously computed $\Sigma$ as an initial guess. 
We use a third-order accurate Runge--Kutta time stepper~\cite{gottlieb1998total}.
For a single species (advected fluid) case, the total number of floating point numbers stored by our scheme is $17 N + o(N)$, where $o(N)$ is the number of grid points.
This includes $5$ state variable arrays (density, energy, and three momenta), which are the \textit{degrees of freedom} of the solution per grid cell.
We also hold $5$ arrays for the Runge--Kutta sub-step, $5$ arrays for the right-hand side of the discretized PDE system, one array for $\Sigma$, and one for the right-hand side of \cref{eqn:igrelliptic}. 
An additional copy of $\Sigma$ is required if Jacobi sweeps are used for the iterative solve.

\subsection{The algorithm}

\begin{algorithm} 
\caption{Compute right-hand side (RHS)}\label{alg}
$(\rho,\rho\bu,E,\alpha) \gets \text{Conservative variables}$ \label{alg:l1} \\
$(\bu,p) \gets \text{Primitive variables}$ \label{alg:l2} \\
$\Sigma \gets \text{Entropic pressure}$ \label{alg:l3} \\
$\mathtt{rhs} \gets \text{Time stepper RHS}$ \label{alg:l4} \\
$\mathtt{vflux} \gets \text{Temp.\ array for viscous flux}$ \label{alg:l5} \\
$\mathtt{coeff} \gets \text{Reconstruction coefficients}$ \label{alg:l6} \\
$\mathtt{igr\_rhs} \gets \text{RHS for elliptic solve}$ \label{alg:l7} \\
$\mathtt{igr\_func()} \gets \text{Routine to compute} \,\, \mathtt{igr\_rhs}$ \label{alg:l8} \\
$\mathtt{viscous()} \gets \text{Routine for viscous flux}$ \label{alg:l9} \\
$\mathtt{flux()} \gets \text{Routine for inviscid flux}$ \label{alg:l10} \\
\For(\tcp*[f]{Loop over time steps}){$t = 1$ \KwTo $T$}{\label{alg:l11}
    \For(\tcp*[f]{Loop over domain}){$\mathtt{dir} \gets \mathtt{(x,y,z)}$}{\label{alg:l12}
        \ForEach{$\mathtt{(i, j, k)}$ $\mathbf{in}$ $\mathtt{cells}$}{\label{alg:l13}
            \For(\tcp*[f]{Reconstruction}){$\mathtt{q \gets -2,3}$}{\label{alg:l14}
                \For{$\mathtt{n \gets (x,y,z)}$}{\label{alg:l15}
                    $\mathtt{\color{seagreen}compute} \,\, \dd_{\mathtt{n}} \bu$ \label{alg:l16} \\
                    $\mathtt{vflux}_\mathrm{L} \gets \mathtt{vflux}_\mathrm{L} 
                        + \mathtt{coeff}_\mathrm{L}\mathtt{(q)}\,\dd_{\mathtt{n}} \bu$
                        \label{alg:l17} \\ 
                    $\mathtt{vflux}_\mathrm{R} \gets 
                        \mathtt{vflux}_\mathrm{R} 
                        + \mathtt{coeff}_\mathrm{R}\mathtt{(q)}\,\dd_{\mathtt{n}} \bu$
                        \label{alg:l18} \\ 
                    \If{$\mathtt{dir = x} \wedge \mathtt{q = 0}$}{\label{alg:l19}
                        $\mathtt{\textcolor{seagreen}{store}}\,\, \dd_{\mathtt{n}} \bu$ \label{alg:l20}
                    } 
                } 
                \If{$\mathtt{dir = x} \wedge \mathtt{q = 0}$}{ \label{alg:l21}
                    $\mathtt{igr\_rhs} \gets \mathtt{igr\_func}\!
                        \left( \dd_x \bu, \dd_y \bu , \dd_z \bu \right)$ \label{alg:l22}
                }               
            }
            \tcp{Recon. density, velocity}
            $\rho_\mathrm{L}, \rho_\mathrm{R} 
                \gets \rho(-2\!:\!3)\ 
                \mathtt{\textcolor{seagreen}{along} \,\, \textcolor{black}{dir}}$
                \label{alg:l23} \\
            $\rho\bu_\mathrm{L}, \rho\bu_\mathrm{R} 
                \gets 
                \rho\bu(-2\!:\!3)\ 
                \mathtt{\textcolor{seagreen}{along} \,\, \textcolor{black}{dir}}$
                \label{alg:l24}\\
            \tcp{Convert to primitive} 
            $\bu_\mathrm{L},\bu_\mathrm{R} \gets 
                \rho\bu_\mathrm{L}, \rho\bu_\mathrm{R}$ \label{alg:l25} \\
            \tcp{Viscous fluxes} 
            $\mathtt{rhs} \gets \mathtt{viscous}(
               \mathtt{vflux}_\mathrm{L}, \mathtt{vflux}_\mathrm{R},
               \bu_\mathrm{L}, \bu_\mathrm{R})$ \label{alg:l26}\\
            \tcp{Recon. remaining variables} 
            $E_\mathrm{L}, E_\mathrm{R} 
                \gets 
                E(-2\!:\!3)\ 
                \mathtt{\textcolor{seagreen}{along} \,\, \textcolor{black}{dir}}$
                \label{alg:l27} \\
            $\alpha_\mathrm{L}, \alpha_\mathrm{R} 
                \gets 
                \alpha(-2\!:\!3)\ 
                \mathtt{\textcolor{seagreen}{along} \,\, \textcolor{black}{dir}}$
                \label{alg:l28} \\
            \tcp{Convert to primitive} 
            $p_\mathrm{L},p_\mathrm{R} \gets 
                E_\mathrm{L},E_\mathrm{R}$ 
                \label{alg:l29} \\
            $\Sigma_\mathrm{L}, \Sigma_\mathrm{R} 
                \gets 0$ \label{alg:l30} \\
            \tcp{IGR contribution in y,z} 
            \If{$\mathtt{dir = y} \ \vee\ \mathtt{dir = z}$}{ \label{alg:l31}
                $\Sigma_\mathrm{L}, \Sigma_\mathrm{R} 
                    \gets 
                    \Sigma(-2\!:\!3)\ 
                    \mathtt{\textcolor{seagreen}{along} \,\, \textcolor{black}{dir}}$ \label{alg:l32} 
            }
            \For(\tcp*[f]{Inviscid fluxes}){
                $d \gets \mathrm{L}, \mathrm{R}$}{ \label{alg:l33}
                    $\mathtt{rhs} \gets \mathtt{flux}
                        (\rho_{\mathtt{d}},\bu_{\mathtt{d}},
                        E_{\mathtt{d}},p_{\mathtt{d}},
                        \alpha_{\mathtt{d}},\sigma_{\mathtt{d}})$
                        \label{alg:l34} 
            }
        }
        \If{$\mathtt{dir \!=\! x}$}{ \label{alg:l35}
            \tcp{IGR elliptic solve} 
            $\Sigma \gets \mathtt{igr\_rhs}$ \label{alg:l36} \\
            \For(\tcp*[f]{IGR x contrib.}){$\mathtt{d} \gets \mathrm{L}, \mathrm{R}$}{ \label{alg:l37}
                    $\rho\bu_\mathrm{L}, \rho\bu_\mathrm{R} 
                        \gets 
                        \rho\bu(-2\!:\!3)\ 
                        \mathtt{\textcolor{seagreen}{along} \,\, \textcolor{black}{dir}}$
                        \label{alg:l38}\\
                    $\bu_\mathrm{L}, \bu_\mathrm{R} \gets 
                        \rho\bu_\mathrm{L}, \rho\bu_\mathrm{R}$ \label{alg:l39} \\
                    $\Sigma_\mathrm{L},\Sigma_\mathrm{R} \gets 
                        \Sigma(-2\!:\!3)\ 
                        \mathtt{\textcolor{seagreen}{along} \,\, \textcolor{black}{dir}}$
                        \label{alg:l40} \\
                    $\mathtt{rhs} 
                        \gets 
                        \mathtt{flux(}\bu_{\mathtt{d}},\sigma_{\mathtt{d}}\mathtt{)}$
                        \label{alg:l41} \\
            }
        }
    }
    $(\rho,\rho\bu,E,\alpha) \gets 
     (\rho,\rho\bu,E,\alpha) +  \mathrm{d}t \cdot \mathtt{rhs}$\label{alg:l42}
}
\end{algorithm}

The key algorithmic kernel of our method computes the right-hand side of the ordinary differential equation obtained from the spatial discretization. 
It is presented in \cref{alg}.
We advance the conservative variables (\cref{alg:l1}) at the cell centers using a 3rd-order accurate Runge--Kutta time stepper, requiring $2$ copies of the state variables.
The flux calculations at the cell boundaries are split dimensionally across the three coordinate directions (\cref{alg:l12}). 
The conservative variables are reconstructed at the cell boundaries using a 5th-order accurate polynomial interpolation scheme (\crefrange{alg:l23}{alg:l28}).

The viscous fluxes require the calculation and reconstruction of velocity gradients (\crefrange{alg:l16}{alg:l18}).
A conversion of the reconstructed conservative variables to their primitive form is performed at the cell boundaries (\cref{alg:l25,alg:l29}).
The Riemann problem at the interface is then solved using a Lax--Friedrichs approximate Riemann solver (\cref{alg:l26,alg:l34}). 
The net flux at the cell center is an input to the time stepper (\cref{alg:l4}) via the right-hand side.
The entropic pressure $\Sigma$ is calculated at the cell centers by solving the elliptic PDE (\cref{alg:l36}) in \cref{eqn:igrelliptic} and incorporated into the right-hand side (\cref{alg:l34,alg:l41}). 
The left side of \cref{eqn:igrelliptic} also requires velocity gradients, which are reused from the viscous flux calculations (\crefrange{alg:l20}{alg:l22}).

\subsection{Optimizations}
 
Our implementation eliminates the storage of the reconstructed states (\crefrange{alg:l23}{alg:l29}), velocity gradients (\crefrange{alg:l16}{alg:l20}), and fluxes (\cref{alg:l26,alg:l34,alg:l41}) and keeps all operations in a single kernel (\cref{alg}). 
The memory footprint is reduced by storing the intermediate variables as thread-local temporary arrays within this kernel.
The algorithm only requires storing $2$ copies of the conservative variables, the net flux at each grid point, and the solution and left side of the elliptic PDE in \cref{eqn:igrelliptic}.

Each thread solves an approximate Riemann problem at the grid cell--cell interface and accumulates its contribution to the right-hand side at overlapping cells via atomic operations to prevent race conditions.
During the reconstruction along the 1st coordinate dimension ($x$, here), the contributions to the left-hand side of equation \cref{eqn:igrelliptic} are computed using the velocity gradients for the viscous fluxes (\cref{alg:l22}). 
The elliptic PDE in \cref{eqn:igrelliptic} is solved after its left side is computed (\cref{alg:l36}). 
The entropic pressure added directly to the flux in the 2nd and 3rd dimensions (\crefrange{alg:l31}{alg:l32}). 
The flux contribution of the entropic pressure in the first dimension is completed separately after the elliptic solve (\crefrange{alg:l36}{alg:l41}).
The above decreases memory use $25$-fold compared to an optimized 5th-order accurate WENO and Riemann solver implementation in the same codebase~\cite{wilfong252}.

\begin{table*}[ht]
    \centering
    \caption{Node and full system properties of the supercomputers tested in this work. 
    TOP500 rankings from June~2025.}
    \begin{tabular}{ r l r r r r l }\toprule
                        & Node Configuration & \# Nodes & Memory [Node, System] & Peak Power & Rmax & TOP500 \\ \midrule
       LLNL El~Capitan & 4 AMD MI300A APU & 11136  & [$\SI{512}{\giga\byte}$, $\SI{5.6}{\peta\byte}$] APU & $\SI{34.8}{\mega\watt}$  & $\SI{1742}{\peta\flops}$ & 1
                        \\[1ex]
       OLCF Frontier    & 4 AMD MI250X GPU & 9472  & [$\SI{512}{\giga\byte}$, $\SI{4.8}{\peta\byte}$] GPU  & $\SI{24.6}{\mega\watt}$  &  $\SI{1353}{\peta\flops}$ & 2     \\
                        & 1 AMD Trento CPU & & [$\SI{512}{\giga\byte}$,  $\SI{4.8}{\peta\byte}$] CPU & & \\[1ex]
       CSCS Alps        & 4 NVIDIA GH200 & 2688 & [$\SI{384}{\giga\byte}$, $\SI{1.0}{\peta\byte}$] GPU  & $\SI{7.1}{\mega\watt}$ & $\SI{435}{\peta\flops}$ & 8\\ 
                        & (4 Grace CPU, 4 Hopper GPU) & & [$\SI{480}{\giga\byte}$, $\SI{1.3}{\peta\byte}$] CPU & &  \\
    \bottomrule
    \end{tabular}
    \label{tab:systems}
\end{table*}

\begin{figure}
    \centering
    % \begin{tikzpicture}[thick]

% \draw[very very thick,seagreen] (0,0) -- (8.75,0)  node[anchor=south east]{CPU};
% \draw[very very thick, darkrust] (0,3) -- (8.75,3) node[anchor=south east]{GPU};
% \draw[very very thick,-latex] (7.7,2.7) -- (8.75,2.7) node[anchor=north east,inner sep=4pt] {Time};

% \draw[*-,] (0.5,1) -- (0.5,0) node[below,align=center] {\color{seagreen} \tt ALLOC \\[-0.1cm] \color{seagreen} $q_{1,2}$};

% \draw[-latex,] (1.5,0) -- (1.5,3) node[above] {\color{darkrust}$q_1$} node[midway,below,rotate=90] {C2C};

% \draw[dashed,-latex,] (3.5,3) -- (3.5,0) node[above,rotate=90, midway] {C2C} node[below]{\color{seagreen}$q_1$};
% \draw[-latex,] (4,0) -- (4,3) node[above] {\color{darkrust}$q_2$} node[midway,below,rotate=90] {C2C};

% \draw[dashed,-latex,] (6,3) -- (6,0) node[above,rotate=90, midway] {$\mathrm{C2C}$} node[below]{\color{seagreen}$q_2$};

% \draw[latex-latex,] (1.5,2.25) -- (3.5,2.25) node[midway,above] {\tt RHS}; 
% \draw[latex-latex,] (4,2.25) -- (6,2.25) node[midway,above] {\tt RHS}; 

% \draw[*-,] (7.7,1) -- (7.7,0) node[below,align=center] {\color{seagreen} $q_3 = f(q_1,q_2)$};

% \end{tikzpicture}

\begin{tikzpicture}

    % CPU Chip
    \draw[draw=none,fill=darkrust,opacity=0.6] (0.05in, 0.15in) rectangle (0.95in,0.85in);
    \draw[draw=none,fill=darkrust,opacity=0.6] (0.15in, 0.05in) rectangle (0.85in,0.95in);
    \draw[draw=none,fill=darkrust] (0.1in, 0.1in) rectangle (0.9in,0.9in);
    \node[anchor=center] at (0.5in, 0.5in) {\textbf{\textcolor{white}{\huge CPU}}};

    \foreach \i in {0.175, 0.225, ..., 0.875} {
        \draw[gray] (0.95in, \i in) -- (1.0in, \i in);
        \draw[gray] (0.00in, \i in) -- (0.05in, \i in);
    }

    \foreach \i in {0.175, 0.225, ..., 0.875} {
        \draw[gray] (\i in, 0.95in) -- (\i in, 1.0in);
        \draw[gray] (\i in, 0.00in) -- (\i in, 0.05in);
    }

    % GPU Chip
    \draw[draw=none,fill=seagreen,opacity=0.6] (2.05in, 0.15in) rectangle (2.95in,0.85in);
    \draw[draw=none,fill=seagreen,opacity=0.6] (2.15in, 0.05in) rectangle (2.85in,0.95in);
    \draw[draw=none,fill=seagreen] (2.1in, 0.1in) rectangle (2.9in,0.9in);
    \node[anchor=center] at (2.5in, 0.5in) {\textbf{\textcolor{white}{\huge GPU}}};

    \foreach \i in {0.175, 0.225, ..., 0.875} {
        \draw[gray] (2.95in, \i in) -- (3.0in, \i in);
        \draw[gray] (2.00in, \i in) -- (2.05in, \i in);
    }

    \foreach \i in {2.175, 2.225, ..., 2.875} {
        \draw[gray] (\i in, 0.95in) -- (\i in, 1.0in);
        \draw[gray] (\i in, 0.00in) -- (\i in, 0.05in);
    }

    % High speed interconnect
    \draw[<->, very thick, line width=1pt, >=latex] (1.1in, 0.3in) -- (1.9in, 0.3in);
    \draw[<->, very thick, line width=1pt, >=latex] (1.1in, 0.4in) -- (1.9in, 0.4in);
    \draw[<->, very thick, line width=1pt, >=latex] (1.1in, 0.5in) -- (1.9in, 0.5in);
    \node[anchor=south, align=center] at (1.5in, 0.55in) {Chip-to-chip \\ interconnect};

    % Chip tails
    \draw[very thick,darkrust] (0.5in, -0.05in) -- (0.5in, -1.825in);
    \draw[very thick,seagreen] (2.5in, -0.05in) -- (2.5in, -1.825in);
    \draw[very thick,->,>=latex] (2.575in, -1.45in) -- (2.575in, -1.825in);
    \node[anchor=south,rotate=270] at (2.6in, -1.6375in) {Time};

    % Memory allocation
    \draw[very thick, darkrust] (0.5in, -0.125in) -- (0.6in, -0.125in);
    \filldraw[darkrust] (0.6in, -0.125in) circle (2pt);
    \node[anchor=west] at (0.65in, -0.125in) {\textcolor{darkrust}{$\texttt{Alloc } q_2$}};

    \draw[very thick, seagreen] (2.4in, -0.125in) -- (2.5in, -0.125in);
    \filldraw[seagreen] (2.4in, -0.125in) circle (2pt);
    \node[anchor=east] at (2.35in, -0.125in) {\textcolor{seagreen}{$\texttt{Alloc } q_1$}};

    % Step 1
    \node[draw,rectangle,gray,very thick,anchor=center,align=center,minimum height=0.375in,minimum width=0.75in] 
        at (1.5in,-0.5in) {$q_2 = q_1$ \\ $q_1 = g(q_1)$}; 
    \draw[very thick,->,>=latex,seagreen] (2.5in, -0.375in) -- (1.875in, -0.375in); 
    % \node[anchor=west, seagreen] at (2.55in, -0.375in) {$q_1$};
    \draw[very thick,->,>=latex,seagreen] (1.875in, -0.625in) -- (2.5in, -0.625in); 
    % \node[anchor=west, seagreen] at (2.55in, -0.625in) {$q_1$};
    \draw[very thick,->,>=latex,darkrust] (1.125in, -0.625in) -- (0.5in, -0.625in); 
    % \node[anchor=east, darkrust] at (0.45in, -0.625in) {$q_2$};

    % Step 2
    \node[draw,rectangle,gray,very thick,anchor=center,align=center,minimum height=0.375in,minimum width=1.125in] 
        at (1.5in,-1.0in) {$q^* = f(q_1)$ \\ $q_1 = g(q_1, q_2, q^*)$}; 
    \draw[very thick,->,>=latex,seagreen] (2.5in, -0.875in) -- (2.0625in, -0.875in); 
    % \node[anchor=west, seagreen] at (2.55in, -0.875in) {$q_1$};
    \draw[very thick,->,>=latex,seagreen] (2.0625in, -1.125in) -- (2.5in, -1.125in); 
    % \node[anchor=west, seagreen] at (2.55in, -1.125in) {$q_1$};
    \draw[very thick,->,>=latex,darkrust] (0.5in, -0.875in) -- (0.9375in, -0.875in); 
    % \node[anchor=east, darkrust] at (0.45in, -0.875in) {$q_2$};

    % Step 3
    \node[draw,rectangle,gray,very thick,anchor=center,align=center,minimum height=0.375in,minimum width=1.125in] 
        at (1.5in,-1.5in) {$q^* = f(q_1)$ \\ $q_1 = g(q_1, q_2, q^*)$}; 
    \draw[very thick,->,>=latex,seagreen] (2.5in, -1.375in) -- (2.0625in, -1.375in); 
    % \node[anchor=west, seagreen] at (2.55in, -1.375in) {$q_1$};
    \draw[very thick,->,>=latex,seagreen] (2.0625in, -1.625in) -- (2.5in, -1.625in); 
    % \node[anchor=west, seagreen] at (2.55in, -1.625in) {$q_1$};
    \draw[very thick,->,>=latex,darkrust] (0.5in, -1.375in) -- (0.9375in, -1.375in); 
    % \node[anchor=east, darkrust] at (0.45in, -1.325in) {$q_2$};

    % C2C Top
    % \draw[->, very thick, >=latex, black] (0.5in, -0.5in) -- (2.5in, -0.5in);
    % \node[anchor=west] at (2.55in, -0.5in) {\textcolor{seagreen}{$q_1$}};
    % \draw[<-, very thick, dashed, >=latex, black] (0.5in, -0.875in) -- (2.5in, -0.875in);
    % \node[anchor=east] at (0.45in, -0.875in) {\textcolor{darkrust}{$q_1$}};
    % \draw[<->, very thick, >=latex, black] (2.25in, -0.5in) -- (2.25in, -0.875in);
    % \node[anchor=north] at (1.5in, -0.5in) {\texttt{C2C}};
    % \node[anchor=south] at (1.5in, -0.875in) {\texttt{C2C}};
    % \node[anchor=south, rotate=270] at (2.25in, -0.6875in) {\texttt{RHS}};

    % C2C Bottom
    % \draw[->, very thick, >=latex, black] (0.5in, -1.0in) -- (2.5in, -1.0in);
    % \node[anchor=west] at (2.55in, -1.0in) {\textcolor{seagreen}{$q_2$}};
    % \draw[<-, very thick, dashed, >=latex, black] (0.5in, -1.375in) -- (2.5in, -1.375in);
    % \node[anchor=east] at (0.45in, -1.375in) {\textcolor{darkrust}{$q_2$}};
    % \draw[<->, very thick, >=latex, black] (2.25in, -1.0in) -- (2.25in, -1.375in);
    % \node[anchor=north] at (1.5in, -1.0in) {\texttt{C2C}};
    % \node[anchor=south] at (1.5in, -1.375in) {\texttt{C2C}};
    % \node[anchor=south, rotate=270] at (2.25in, -1.1875in) {\texttt{RHS}};

    % Computation
    % \draw[very thick, darkrust] (0.5in, -1.625in) -- (1in, -1.625in);
    % \filldraw[darkrust] (1in, -1.625in) circle (2pt);
    % \node[anchor=west] at (1.05in, -1.625in) {\textcolor{darkrust}{$q_3 = f(q_1, q_2)$}};

\end{tikzpicture}
    \Description{Schematic showing where memory is stored and when it is used with unified memory enabled.}
    \caption{
        Schematic of the chip-to-chip (C2C) transfers of intermediate time-step variables between the on-node CPU and GPU devices.
        The time sub-steps are $q_{1,2}$ and the full step integration is stored in $q_1$.
    }
    \label{fig:unified}
\end{figure}

\subsection{Unified Memory}\label{sec:unified}

The unified memory approach uses unified-shared-memory (USM) mode on the MI300A (El~Capitan), AMD InfinityFabric for CPU--GPU connections on Frontier, and the NVLink~C2C connection on Grace~Hopper (Alps),
\Cref{fig:unified} shows this strategy, using the full capacity of the compute node while expanding beyond the GPU memory without incurring a meaningful performance penalty.
This allows us to store the intermediate Runge-Kutta stage on the host, reducing GPU memory use by up to a factor of $12/17$.

\subsubsection{AMD approach for El~Capitan}

On the AMD~MI300A, the CPU and GPU share a single physical HBM pool.
We compile MFC with OpenMP offloading directives using AMD's next-generation Flang compiler and OpenMP's USM mode.
All variables have a single copy in memory and are accessible from CPUs and accelerators with no data movement.
CCE's OpenACC implementation is also used for speed comparisons.

\subsubsection{AMD approach for Frontier}

Our approach to Frontier involves using OpenACC or OpenMP, as well as HPE's CCE or the AMD~Flang next-generation compiler, for performance comparisons between compilers and unified memory implementations.

None of the vendor compilers available on Frontier support an equivalent of USM for OpenACC, though both CCE and AMD's ROCm compiler support OpenMP's USM mode.
CCE allows users to allocate device-accessible memory and request OpenACC to omit a separate device copy, effectively eliding the need for USM via UVM.
The AMD heterogeneous system architecture (HSA) requires that GPU memory be accessible by the host.
To reduce HBM use on Frontier, we allocate a single time-step stage as device-resident memory (via \lstinline{hipMalloc}) and the second as pinned host memory (via \lstinline{hipMallocManaged} and \lstinline{hipMemAdvise}).
With CCE and OpenACC, we set \lstinline{CRAY_ACC_USE_UNIFIED_MEM=1}.
Hence, the CCE OpenACC runtime detects that these arrays do not need to be mapped and instead uses zero-copy across the AMD Trento--MI250X InfinityFabric.

With OpenMP, the mapping of the Fortran pointers associated with the device-resident memory within the MFC data structures is omitted.
No redundant memory copy is created due to MP's treatment of the derived type mapping with Fortran pointer components.
MFC can also run on Frontier in USM mode, but was found to be less performant on Frontier than a UVM strategy discussed in \cref{sec:alpsunified} below, and is not used in this work.
In summary, we reduce memory usage on both the host and device without compromising performance.

With CCE, the buffers passed to MPI are mapped to device memory by the OpenACC runtime.
We use GPU-aware~MPI to communicate GPU buffers directly.
This work also tests a development build of AMD's new Flang compiler that requires a beta version of the ROCm~7 runtime.
API changes in this runtime compared to ROCm~6 require us to disable GPU-aware~MPI for AMD~Flang builds.

\subsubsection{NVIDIA approach for Alps}\label{sec:alpsunified}

The Alps nodes have a hardware-coherent system that couples a Grace CPU and a Hopper GPU using a \SI{900}{\giga\byte\per\second} NVLink-C2C connection.
This allows both processors to access all system memory at high speeds coherently and consistently. 
In addition to the usual CUDA allocators, GPU memory can be allocated using system allocators such as \lstinline{malloc}.
The unified memory concept maps to the Grace~Hopper superchip and is used herein.

To realize out-of-core GPU computations, we compile and link via \lstinline{-gpu=mem:unified}, instructing the compiler to use CUDA Unified Memory.
This provides a single unified address space for the CPU and GPU.
Our optimizations leverage this infrastructure via a \emph{zero-copy} strategy, where the most frequently accessed data are hosted in GPU memory and the least frequently accessed are in CPU memory.
We avoid data movement during the simulation and only perform local or remote direct accesses.
This eliminates the duplication of host and device buffers, thereby maximizing the simulation size.
To fine-tune data placement, we provide memory hints to the CUDA driver via \lstinline{cudaMemAdvise} and \lstinline{cudaMemPrefetchAsync}.
For the buffers that stay in CPU memory for the full lifetime of the simulation, we use \lstinline{pinned} allocations.
This results in a minimally intrusive out-of-core implementation that does not sacrifice GPU performance.

The time step updates in the Runge--Kutta scheme are rearranged so that only the current sub-step is passed to the right-hand side routine.
The buffer holding the previous state is used to update the current Runge--Kutta state.
With this rearrangement, we only store one sub-step and the right-hand side buffer in GPU memory.
Thus, the intermediary sub-step is always in CPU memory, freeing GPU memory and increasing simulation size without sacrificing performance.
The time step updates access Runge--Kutta sub-steps from their physical locations via zero-copy, allowing simultaneous access to data from both GPU and CPU memory through the C2C NVLink.
We include flexibility in hosting the IGR temporary variables in either GPU or CPU memory, further reducing the memory footprint from $12/17$ to $10/17$ and scaling to larger problems with little performance impact.

We communicate via CUDA-aware MPI and GPUDirect, avoiding the need for staging in host memory.
We allocate the send and receive buffers on the GPU using OpenACC capture to guide MPI in selecting the GPU path. 
This strategy enables separate memory for the send/receive halo buffers.

\begin{figure}
    \centering
    \begin{subfigure}{0.45\columnwidth}
        \includegraphics[width=\textwidth]{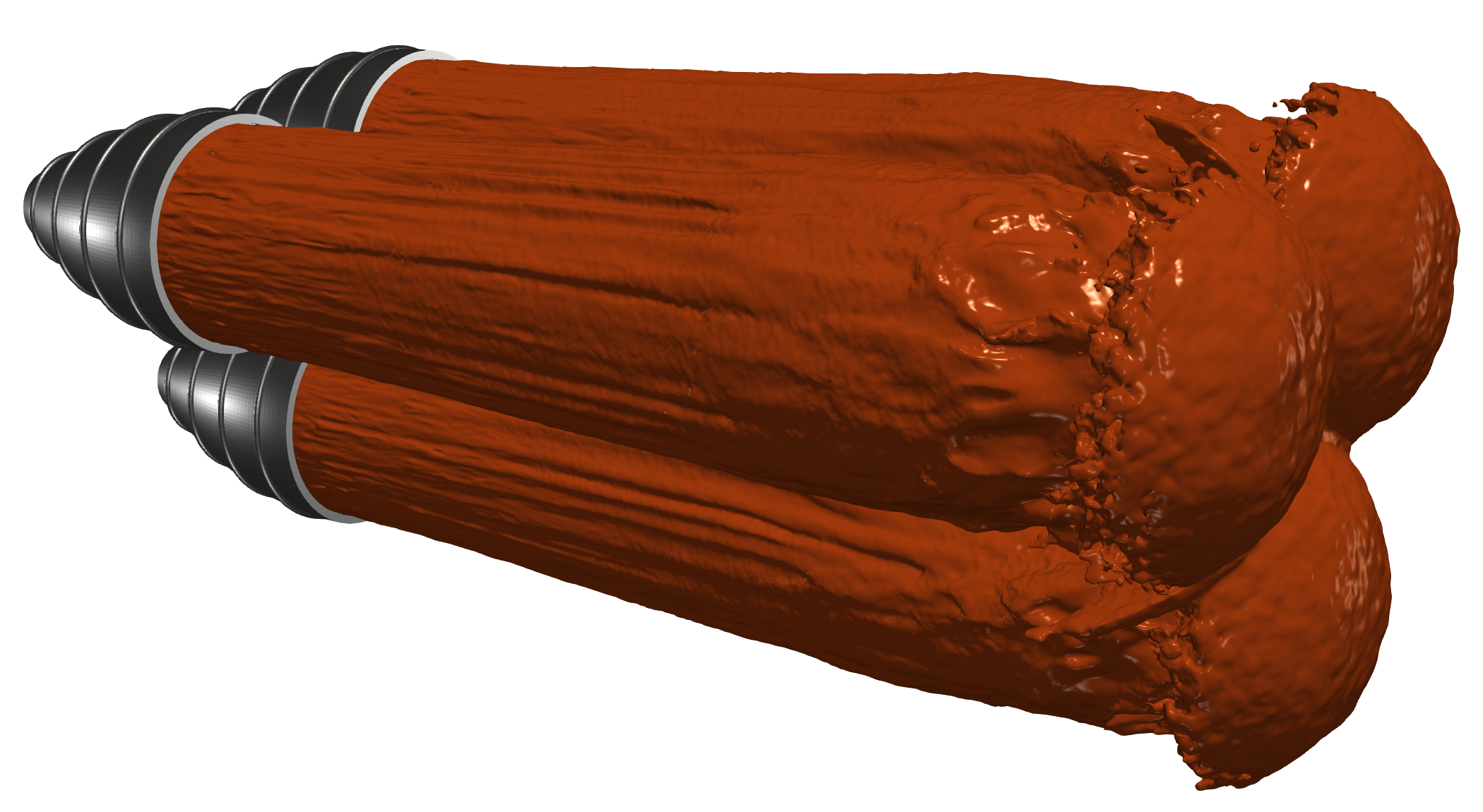}
        \caption{FP16/FP32}
    \end{subfigure}   
    \begin{subfigure}{0.45\columnwidth}
        \includegraphics[width=\textwidth]{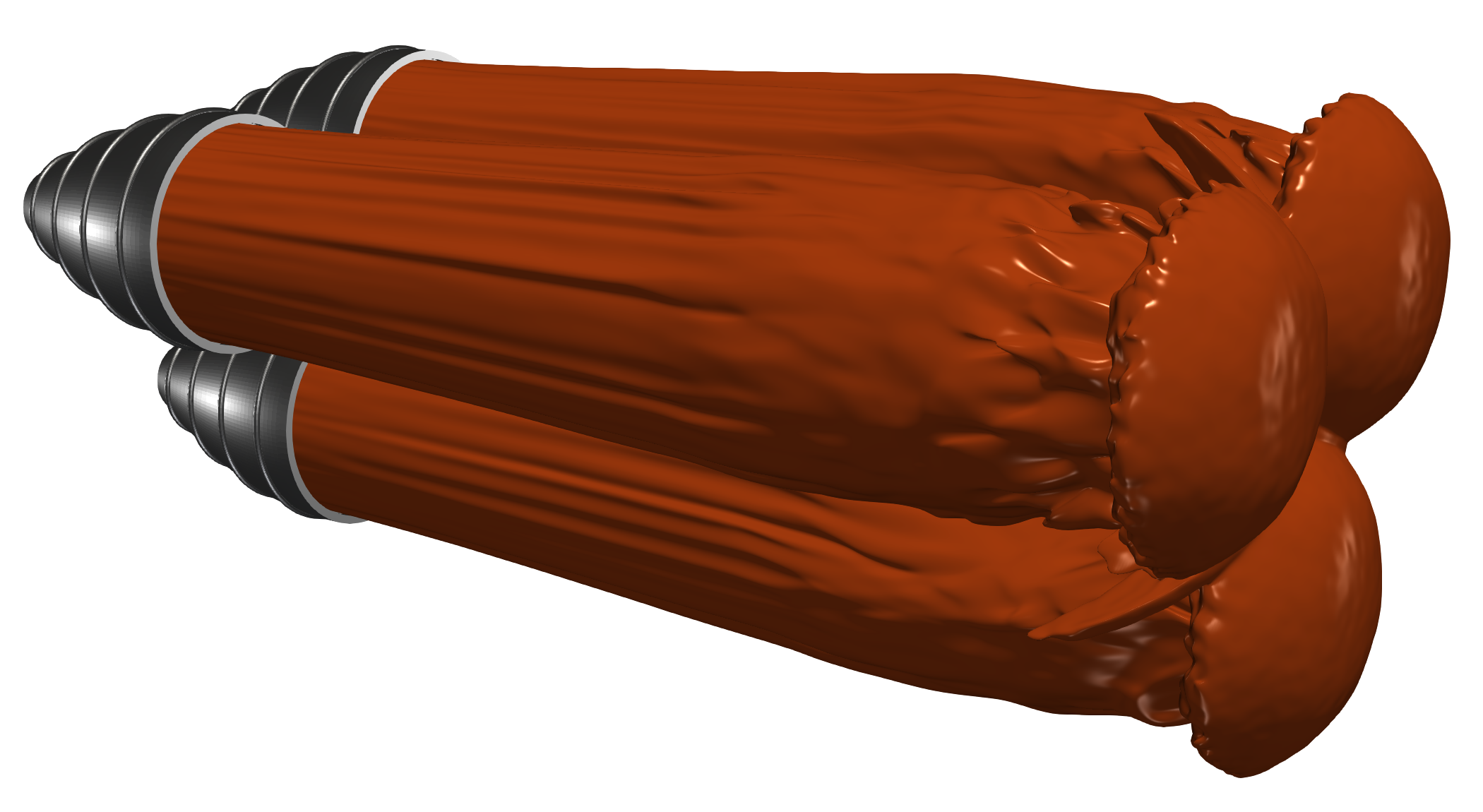}
        \caption{FP32}
    \end{subfigure}   
    \begin{subfigure}{0.45\columnwidth}
        \includegraphics[width=\textwidth]{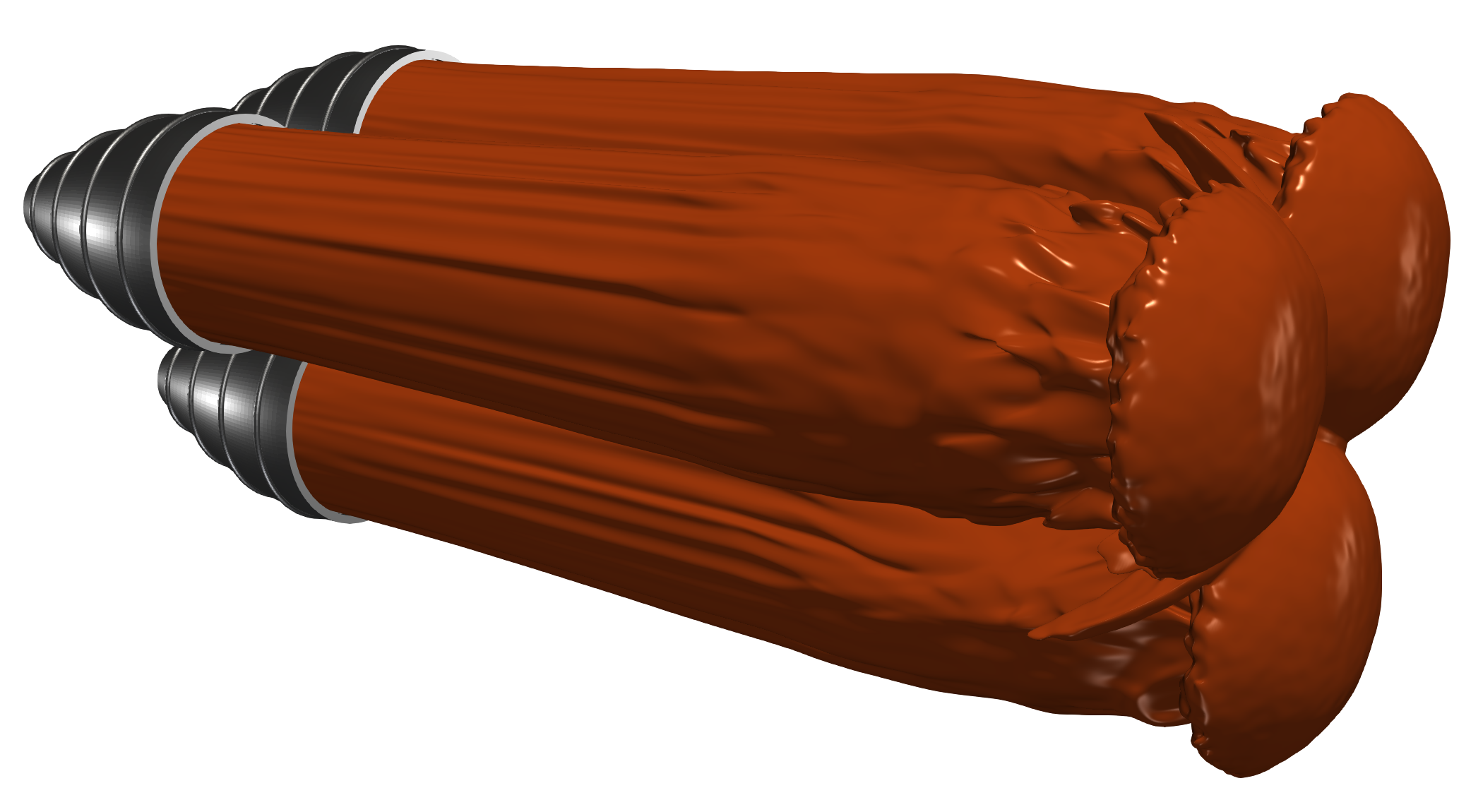}
        \caption{FP64}
    \end{subfigure} 
    \begin{subfigure}{0.45\columnwidth}
        \includegraphics[width=\textwidth]{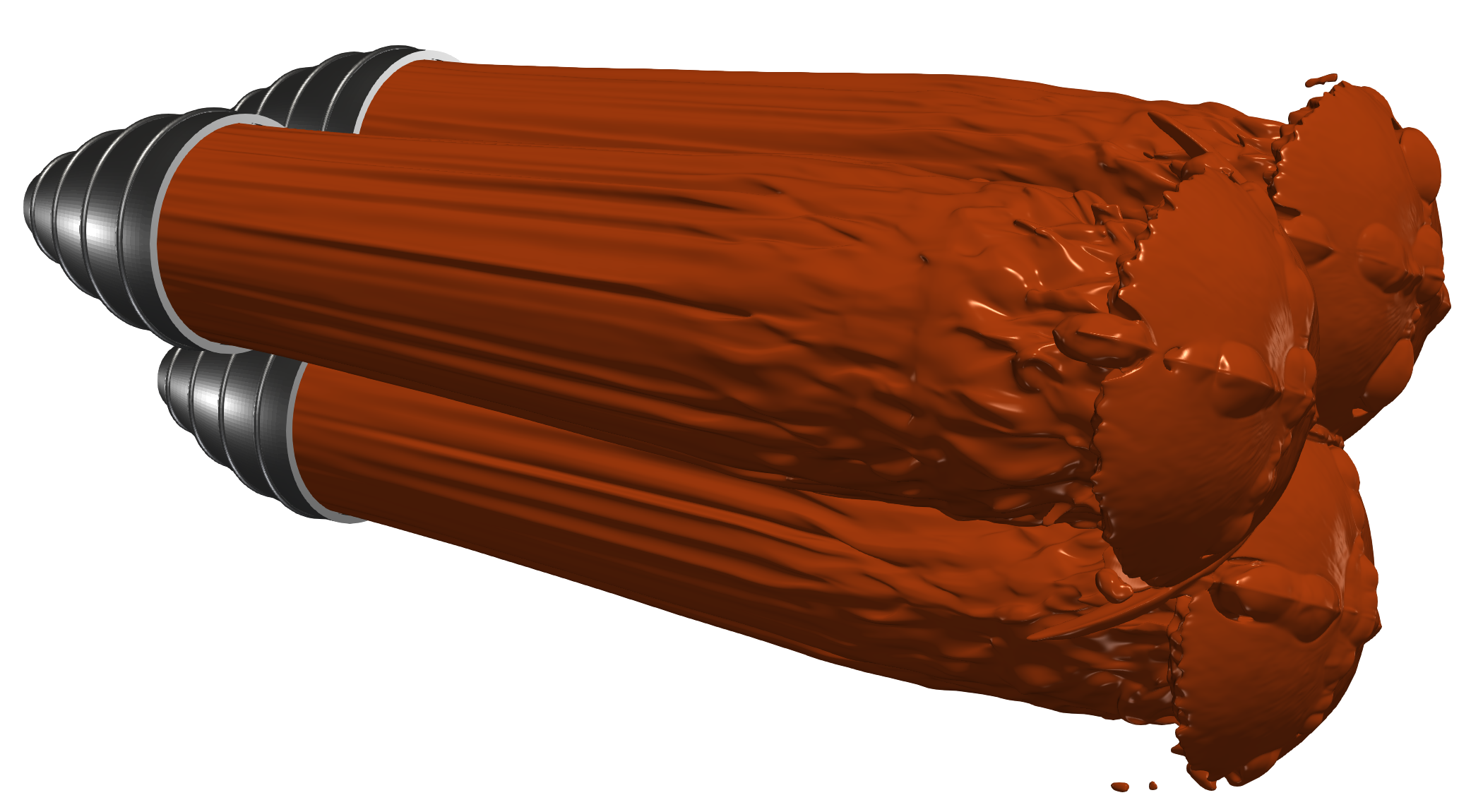}
        \caption{FP64 (Baseline)}
    \end{subfigure}
    \Description{Comparison of simulation results for a three jet configuration with different working precisions.}
    \caption{
    Visualization of a three-engine configuration and its plumes using (a) FP16, (b) FP32, (c) FP64 storage, and (d) the baseline numerics.
    The contours indicate where the velocity exceeds the free stream flow, and the initial state is seeded with smooth, random noise in all cases.
    FP32 and FP64 yield visually indistinguishable results.
    Visual differences in the FP16 case are solely due to the earlier onset of physical flow instabilities, yet they remain faithfully representative of the flow features.
    The grid-dependent nature of the baseline shock-capturing approach results in spurious grid-alignment artifacts.
    }
    \label{fig:precisionComp}
\end{figure}

\subsection{Mixed Precision}

On all systems, we implement a FP16/32 mixed-precision strategy; computations are performed in FP32 and stored in FP16.
On Alps we use NVHPC and on AMD systems the AMD~Flang compiler, both of which support FP16 natively.

This strategy is numerically viable for various compressible flow simulations, as discussed in \cref{sec:mixed}.
This strategy further doubles the maximum simulation size compared to state-of-the-art shock capturing methods we are comparing against, which are numerically unstable even at single precision.

While most directive constructs are readily applicable to half-precision, support for atomic updates in the NVHPC compiler was just added to its SDK and is used here.
Here, atomic updates are more performant in their vectorized version, which we implement in a batched manner.
Using FP16/32, we enable GH200 simulations \emph{exceeding 100T grid points} by extrapolating our results from CSCS~Alps to JSC~JUPITER, which has the same primary architecture.

\Cref{fig:precisionComp} compares simulation results for engine plumes performed with FP16, FP32, and FP64 storage with the FP64 baseline numerics.
Double and single precision results are visually indistinguishable.
Half-precision storage yields visually different results.
They arise from the more rapid onset of the hydrodynamic instabilities due to their seeding with numerical noise.
At longer simulation times, we expect all precision results to be equivalent.
Grid-aligned artifacts appear in the baseline numerics result due to the grid-dependent nature of the shock-capturing approach.

\section{How Performance was Measured}

\subsection{HPC platforms}

We performed calculations on LLNL~El~Capitan, OLCF~Frontier, and CSCS~Alps.
\Cref{tab:systems} shows the performance attributes of these systems.

\subsubsection{LLNL El~Capitan}

The El~Capitan nodes have four AMD MI300A APUs. Each APU combines 24 EPYC Zen~4 CPU cores and 228 CDNA~3 compute units with a single \SI{128}{\giga\byte} layer of HBM3 memory for both processor types.
Nodes are interconnected via HPE Cray Slingshot-11 Ethernet in a dragonfly topology with four \SI{200}{\giga\byte\per\second} NICs per node.
The GPUs can be programmed with unified shared memory (USM) to avoid duplicate host--device memory addresses.
We use this strategy via OpenMP target offload (see \cref{sec:unified}).

\subsubsection{OLCF Frontier}
Each Frontier node has one 64-core AMD EPYC Trento CPU and four AMD~MI250X GPUs.
Each MI250X GPU contains two Graphics Compute Dies (GCDs) with \SI{64}{\giga\byte} of HBM2E each (or \SI{128}{\giga\byte} per GPU). 
The total system memory is \SI{9.6}{\peta\byte}, equally split between HBM2E~GPU and DDR4~CPU memory.
Frontier uses HPE Slingshot Ethernet for interconnects with a 3-hop Dragonfly topology and four \SI{200}{\giga\byte\per\second} NICs per node, which are attached to the GPUs.

\subsubsection{CSCS Alps}

Alps nodes have 4~NVIDIA Grace~Hopper GH200 superchips. 
Each GH200 has a Hopper GPU with \SI{4}{\tera\byte\per\second} of memory bandwidth for its \SI{96}{\giga\byte} of HBM3 memory and a Grace CPU with 72~Arm-v9 cores with \SI{500}{\giga\byte\per\second} bandwidth for its \SI{120}{\giga\byte} of LPDDR5 memory. 
NVLink-C2C connects the CPU and GPU, enabling efficient data movement.
We use this new capability herein (see \cref{sec:unified}).
Alps's nodes interconnect through HPE Slingshot with \SI{200}{\giga\byte\per\second} injection bandwidth per superchip.

\subsection{Software environment and performance baseline}

We base our implementation on MFC~\cite{bryngelson19_cpc,wilfong252}\footnote{Openly available at \url{https://github.com/MFlowCode/MFC}}, a compressible flow solver that implements state-of-the-art numerical shock capturing, as well as the IGR implementation of this work.
All of MFC's schemes scale ideally to 100\% of LLNL El~Capitan, OLCF Frontier, and CSCS~Alps, among other previous and existing flagship supercomputers~\cite{radhakrishnan24,elwasif23}.

MFC offloads computation onto GPU and superchip/APU devices via either OpenMP or OpenACC, and uses metaprogramming to abstract away and automate vendor-specific or otherwise burdensome optimizations.
MFC has a history of being used to simulate compressible multi-species, phase, and chemically reacting fluid flows~\cite{charalampopoulos21,bryngelson19_whales,bryngelson23,cisneros25}.
Performance results are measured using a representative three-dimensional simulation of the exhaust plume of a single Mach~10 jet. 

We use MFC's optimized implementation of WENO nonlinear reconstructions and HLLC approximate Riemann solves as a baseline for performance comparisons~\cite{wilfong252}.
This implementation matches or outperforms other codebases with the same and similar reconstruction schemes.
On Frontier and El~Capitan we test HPE's CCE~19.0.0 compiler and a pre-release of AMD's Next Generation Flang compiler with HPE~Cray MPICH-8.1.31 for messaging.
For both compilers, we specify \lstinline{-O3} for optimization.
With CCE, we also apply the \lstinline{-haggress} flag to improve performance in large kernels.
On Alps, we use a pre-release of the NVIDIA HPC SDK~25.9 and HPE Cray MPICH~8.1.30 for messaging.
We also use the compile-time flag \lstinline{-gpu=fastmath}, which yields a performance improvement for single-precision computation on NVIDIA devices.
The latest NVIDIA HPC SDK exhibits an FP64 performance regression; we use NVHPC~SDK~24.3 instead.

\subsection{Measurement tools}\label{sec:measurement}

We measure execution time using application internal timers, including the standard \lstinline[language=SQL]{cpu_time} and \lstinline[language=SQL]{system_clock} procedures.
On Frontier and El~Capitan, we periodically sample the instantaneous GPU or APU power draw by reading the virtual file system for the AMD driver.
The power reading is the same as reported by \lstinline{rocm-smi}, but reading from the virtual filesystem has lower overhead.
On Frontier, \lstinline{rocm-smi} collects GPU and HBM power only.
On El~Capitan \lstinline{rocm-smi} includes CPU, GPU, and memory power.
On Alps, \lstinline{nvidia-smi} records the module (CPU, GPU, and memory) power draw.

On all platforms, we post-process the results to account only for power draw during time-stepping, which is then averaged and multiplied by the average time per time step.
The number of grid points further normalizes energy use. 

\section{Performance Results}\label{sec:performanceresults}

\begin{table}[t]
    \centering
    \caption{
    Wall time for representative simulations, quantified via nanoseconds per grid cell per time step.
    The baseline method is compared to the current work.
    For the AMD~MI300A, USM mode is used, so no in-core quantities are presented; other unified cases use UVM.
    AMD device cases were run using Cray~CCE~19.0.0 and AMD~Flang~Preview~7.0.5 compilers, using OpenACC and OpenMP offloading, with the best results presented.
    AMD~Flang is the only current compiler option for native FP16/32 on all AMD GPU/APU devices.
    NVHPC is used for the GH200 cases.
    }
    \begin{tabular}{r r r r c } \toprule
        Device         & \makecell[r]{Baseline\\ (in-core)} & \makecell[r]{IGR\\(in-core)} & \makecell[r]{IGR\\(unified)} & \\ 
        \midrule
         GH200          & 16.89 & 3.83  & 4.18  & \multirow{3}{*}{\rotatebox[origin=r]{-90}{\bf \;\,FP64}}   \\
        MI250X GCD     & 69.72 & 13.01 & 19.81 &   \\  %cray
        MI300A          & 29.50 & $^\dagger$--- & 7.21 & \\ % cray
        \midrule
        GH200          & $^\ast$N/A & 2.70  & 2.81  & \multirow{3}{*}{\rotatebox[origin=r]{-90}{\bf \;\;FP32}} \\
        MI250X GCD     & $^\ast$N/A & 9.12 & 13.03 & \\ %cray
        MI300A        & $^\ast$N/A & $^\dagger$---  & 4.19 & \\ %cray
        \midrule
        GH200 &  $^\ast$N/A & 3.06 & 3.07 & \multirow{3}{*}{\rotatebox[origin=r]{-90}{\bf FP16/32}} \\ 
        MI250X GCD & $^\ast$N/A & 22.63 & 24.71 & \\ %amdflang
        MI300A     &  $^\ast$N/A & $^\dagger$---  & 17.39 & \\ %amdflang
        \bottomrule
    \end{tabular} \\[1ex]
    {\footnotesize $^\ast$Numerically unstable; $^\dagger$MI300A is always unified}
    \label{tab:grindTimes}
\end{table}

\begin{figure}[ht]
    \centering
    \tikzsetnextfilename{weakScaling}
    \begin{tikzpicture}
    \begin{groupplot}[
      group style={
        group size=1 by 3,
        vertical sep=1.5em,
      },
    ]
    \nextgroupplot[
        xmode = log, 
        xticklabel=\empty,
        width = 3in,
        height = 0.625in, 
        xmin=48,
        xmax=1e5,
        ymin = 0, 
        ymax = 1.5,
        legend style={legend cell align=left,at={(0.5,1.0)},
            anchor=south,draw=none,fill=none,
            inner sep=5pt,legend columns=3,
            /tikz/every even column/.append style={column sep=0.3cm}},
        ]
        \begin{scope}[on background layer]
            \draw[draw=none,fill=black!10]
                (axis cs:44544,0) rectangle (axis cs:1e5,1.5);
        \end{scope}
            
        \addplot[black!70,dashed,very thick,domain=24:44544] {1.0};
        \addlegendimage{only marks,mark size=2.5};
        \addlegendimage{
            area legend,
            fill=black!10,
            draw=none,
            legend image code/.code={
                \path[fill=black!10] (0cm,-0.1cm) rectangle (0.6cm,0.15cm);
                \draw[very thick,black] (0cm,-0.1cm) -- (0.0cm,0.15cm);
            },
        }
        \addplot[seagreen,thick,mark=*,mark size=2.5,only marks]
            coordinates {
                (64,1.0) (192,0.9975) (512,1.00071) (1536,0.9979) (4096,0.9975) (12294,1.02233) (43000,1.03107)
            };
        \draw[thick] (axis cs:44544,0) -- (axis cs:44544,1.5);
        \node at (axis cs:48, 0) [above right] {(a) El Capitan (MI300A)};
        \legend{Ideal,Measured,Full System}
    \nextgroupplot[
        xmode = log, 
        ylabel = {Normalized Wall Time},
        xticklabel=\empty,
        width = 3in,
        height = 0.625in, 
        xmin=48,
        xmax=1e5,
        ymin = 0, 
        ymax = 1.5,
    ]
        \begin{scope}[on background layer]
            \draw[draw=none,fill=black!10]
                (axis cs:37888,0) rectangle (axis cs:1e5,1.5);
        \end{scope}
            
        \addplot[black!70,dashed,very thick,domain=24:37888] {1.0};
        \addplot[rust,thick,mark=*,mark size=2.5,only marks]
            coordinates {
                (64,1.0) (192,0.998) (512,1.0129) (1536,0.997) (4096,1.0000) (12288,0.998) (37632,0.998)
            };
        \draw[thick] (axis cs:37888,0) -- (axis cs:37888,1.5);
        \node at (axis cs:48, 0) [above right] {(b) Frontier (Trento+MI250X)};
    \nextgroupplot[
        xmode = log, 
        xlabel = {Number of Devices},
        width = 3in,
        height = 0.625in, 
        xmin=48,
        xmax=1e5,
        ymin = 0, 
        ymax = 1.5,
        ]
        \begin{scope}[on background layer]
            \draw[fill=black!10,draw=none] 
                (axis cs:10752,0) rectangle (axis cs:1e5,1.5); 
        \end{scope}
 
        \addplot[black!70,dashed,very thick,domain=24:10752,forget plot] {1.0};
        \addplot[purp,thick,mark=*,mark size=2.5,only marks] 
            coordinates {
                (64,1.0) (192,1.00064) (512,1.00319) (1536,1.00206) (4096, 1.00142) (9200,1.00013) 
            };
        \draw[thick] (axis cs:10752,0) -- (axis cs:10752,3);
        \node at (axis cs:48, 0) [above right] {(c) Alps (GH200)};

    \end{groupplot}
    
\end{tikzpicture}
    \Description{Plots showing ideal weak scaling on LLNL El~Capitan, OLCF Frontier, and CSCS Alps.}
    \caption{
    Weak scaling performance for a representative thruster problem on LLNL El~Capitan (number of MI300As), OLCF Frontier (number of MI250Xs), and CSCS Alps (number of GH200s).
    The full system is indicated for each case.
    All runs use unified memory, FP16/32 mixed precision, and a 16~node configuration for scaling comparison, which ensures that all MPI communication directions are touched.
    On El~Capitan, we see $\bm{97\%}$ efficiency out to 10750~MI300As.
    On Alps, ideal ($\bm{\approx 100\%}$) scaling is observed at 9.2K~GH200s.
    On Frontier, we also achieve ideal ($\bm{\approx 100\%}$) efficiency at 37.6K~MI250X GPUs, with a maximum problem size of 200T grid points.
    }
    \label{fig:weakScaling}
\end{figure}

\begin{figure}[ht]
    \centering
    \tikzsetnextfilename{strongScaling}
    \begin{tikzpicture}
    \begin{groupplot}[
        group style={
            group size=1 by 3,
            vertical sep=1.5em,
        },
    ]
    \nextgroupplot[
        xmode=log,
        ymode=log,
        xtick={8,16,32,64,128,256,512,1024,2048,4096,8192,16384},
        xticklabel=\empty,
        ytick={1, 4, 16, 64, 256, 1024},
        log basis x={2},
        log basis y={2},
        ylabel={Speedup},
        width=2.75in,
        height=0.625in,
        xmin=8,
        xmax=2^14,
        ymin=1,
        ymax=2^11,
        legend style={legend cell align=left,at={(0.5,1.0)},
            anchor=south,draw=none,fill=none,
            inner sep=5pt,legend columns=4,
            /tikz/every even column/.append style={column sep=0.2cm},
        },
        ]
        
        \begin{scope}[on background layer]
            \draw[fill=black!10,draw=none] 
                (axis cs:11136,1) rectangle (axis cs:2^14,2^11); 
            \addplot +[mark=none,very thick,black,forget plot] coordinates {(11136, 1) (11136, 2048)};
        \end{scope}

        % Ideal
        \addplot[black,dashed,very thick,domain=8:11136, samples=10,forget plot] {x/8};
        \addlegendimage{black,dashed,very thick}
        \addlegendentry{Ideal} 
        
        \addlegendimage{%
            area legend,%
            fill=black!10,%
            draw=none,%
            legend image code/.code={%
              \path[fill=black!10] (0cm,-0.1cm) rectangle (0.6cm,0.15cm);
              \draw[very thick,black] (0cm,-0.1cm) -- (0.0cm,0.15cm);
            }
        }
        \addlegendentry{Full System};

        % Frontier mixed
        \addplot[thick,mark=triangle*,mark size=2.75pt,color=purp] coordinates {
            (8, 1.0)
            (16, 1.95976)
            (48, 5.45906)
            (128, 14.7395)
            (384, 42.4318)
            (1024, 106.08)
            (3072, 262.958)
            (10000, 549.657)
        };
        \addlegendentry{USM};

        \addlegendimage{thick,mark=*,darkrust};
        % \addlegendimage{thick,mark=*,seagreen};
        \addlegendentry{UVM};
        % \addlegendentry{No UVM};
        
        \node at (axis cs:8,2048) [below right] {(a) El Capitan (MI300A)};
    
    \nextgroupplot[
        xmode=log,
        ymode=log,
        xtick={8,16,32,64,128,256,512,1024,2048,4096,8192,16384},
        xticklabel=\empty,
        ytick={1, 4, 16, 64, 256, 1024},
        log basis x={2},
        log basis y={2},
        ylabel={Speedup},
        width=2.75in,
        height=0.625in,
        xmin=8,
        xmax=2^14,
        ymin=1,
        ymax=2^11,
        ]
        
        \begin{scope}[on background layer]
            \draw[fill=black!10,draw=none] 
                (axis cs:9450,1) rectangle (axis cs:2^14,2^11); 
            \addplot +[mark=none,very thick,black,forget plot] coordinates {(9450, 1) (9450, 2^11)};
        \end{scope}

        % Ideal
        \addplot[black,dashed,very thick, domain=8:9450, samples=10] {x/8};

        % Frontier mixed
        \addplot[darkrust,thick,mark=*] 
            table[col sep=comma,header=true,x index=0,y index=4] 
            {figures/strongScaling/Strong_Scaling_Frontier_mixed.csv};
        % \addplot[seagreen,thick,mark=*] 
        %     table[col sep=comma,header=true,x index=0,y index=3]
        %     {figures/strongScaling/Strong_Scaling_Frontier_mixed.csv};

        \node at (axis cs:8,2048) [below right] {(b) Frontier (Trento+MI250X)};
 
    \nextgroupplot[
        xmode=log,
        ymode=log,
        xtick={8,16,32,64,128,256,512,1024,2048,4096,8192,16384},
        ytick={1, 4, 16, 64, 256, 1024},
        log basis x={2},
        log basis y={2},
        ylabel={Speedup},
        xlabel={Number of Nodes},
        width=2.75in,
        height=0.625in,
        xmin=8,
        xmax=2^14,
        ymin=1,
        ymax=2^11,
        ]
        \begin{scope}[on background layer]
            \draw[fill=black!10,draw=none] 
                (axis cs:2688,1) rectangle (axis cs:2^14,2^11); 
            \addplot +[mark=none,very thick,black] coordinates {(2688, 1) (2688, 2^11)};
        \end{scope}

        % Ideal
        \addplot[black,dashed,very thick, domain=8:2688, samples=10] {x/8};

        % Alps Mixed
        \addplot[darkrust,thick,mark=*] table[col sep=comma,header=false,x index=0,y index=3] {figures/strongScaling/alpsStrongMixed.csv};
        % \addplot[seagreen,thick,mark=*] table[col sep=comma,header=false,x index=0,y index=5] {figures/strongScaling/alpsStrongMixed.csv};

        \node at (axis cs:8,2048) [below right] {(c) Alps (GH200)};
    \end{groupplot}
\end{tikzpicture}
    \Description{Plots showing efficient strong scaling on LLNL El~Capitan, OLCF Frontier, and CSCS Alps.}
    \caption{
        Strong scaling on all systems for the same problem configuration as \cref{fig:weakScaling}.
        All results are for FP16/32 mixed precision, though FP32 and FP64 show similar results (FP32 shown in \cref{fig:strongScalingcomparison}).
        The full system is shown for each case.
        Following \cref{fig:weakScaling}, we base all speedups on an 8~node configuration, such that all communication directions are touched.
    }
    \label{fig:strongScaling}
\end{figure}

\subsection{Time step cost}\label{sec:timestepcost}

\Cref{tab:grindTimes} shows the normalized grind times for a problem solved using WENO reconstructions and HLLC approximate Riemann solves with in-core computation (current state of the art, optimized implementation in MFC~\cite{wilfong252}), IGR with in-core computation, and IGR with unified memory on one GH200 on Alps, one MI250X GCD on Frontier, and one MI300A APU on El~Capitan.
The grind time is defined as nanoseconds per grid cell per time step, used to normalize against the different problem sizes that fit within device memory.
Smaller grind times indicate shorter time to solution.

GH200 metrics are collected using separate memory mode for in-core cases and unified memory mode for out-of-core/unified cases.
Performance impacts of less than 5\% are observed when moving from in-core to unified computation on the GH200 architecture.
We used the NVHPC~24.3 compiler for the GH200 FP64 cases due to performance regressions in FP64 performance in newer versions. 
However, an unrelated regression in NVHPC~24.3 was observed when switching from separate memory to unified memory in-core computation, resulting in a $9\%$ performance hit in that case (see \cref{tab:grindTimes}).
The performance difference between in- and out-of-core unified memory builds is less than $5\%$.

When transitioning from in-core computation to unified memory computation on Frontier, performance degradation of $42\%$ and $51\%$ is observed for FP32 and FP64, owing to the slower xGMI links between the Trento CPU and MI250X GCDs (\SI{72}{\giga\byte\per\second} each) compared to the C2C bandwidth between the Grace~CPU and Hopper~GPU (\SI{900}{\giga\byte\per\second}). 
On El~Capitan, there is no difference between unified/in-core as the device shares a unified HBM pool.

The increased grind time in calculations that use unified memory to increase problem size per device on Frontier and Alps results from the exchange of conservative variable buffers between the CPU and GPU at each Runge--Kutta update.
On all devices, the time to solution is reduced by a factor of approximately 4 when comparing WENO to IGR in FP64.
FP64 is broadly recommended for ENO-type shock-capturing due to catastrophic cancellation~\cite {Brogi2024OpenFOAMPrecision}. Our approach can even handle even mixed FP16/FP32 precision.
This reduces the time to solution by a factor of at least 6 compared to the baseline.

For FP16/32, we observe a performance regression on all devices compared to FP32, but we expect to exceed or match FP32 performance upon further code and compiler releases and optimizations.
We use the AMD~Flang beta compiler and a pre-release version of the NVHPC~SDK~25.9.  
Further compiler improvements are staged for non-beta AMD~Flang releases, and the performance hit can be expected to be similar to the NVHPC strategy.

\subsection{Scaling}

\Cref{fig:weakScaling} shows the weak scaling performance on LLNL El~Capitan, OLCF Frontier, and CSCS Alps.
Unified memory and FP16/32 mixed precision are used on all systems.
A weak scaling efficiency of 97\% is observed when scaling from 64 to 43K~MI300As on El~Capitan. 
On Frontier, we achieve perfect (100\%, up to measurement variance) weak scaling when scaling from 64 to 37.6K~MI250X GPUs (128 to 75.2K~GCDs).
Likewise, on CSCS Alps, we achieve perfect weak scaling when scaling from 64 to 9.2K~GH200s. 
Better than $95\%$ efficiencies are also seen for FP32 and FP64 precision (not~shown).

We exceed \emph{200T grid cells}, or 1~quadrillion degrees of freedom (5~state variables) on Frontier using 37.6K~MI250X GPUs (9408~nodes), accommodating $1386^3$ grid points per GCD with UVM, and FP16/32 mixed precision.
On Alps, a problem size of $1611^3$ is used per GH200 and UVM with FP16/32 mixed precision, amounting to 45T grid point simulation on the full system (2688~nodes).
On the recently deployed JSC~JUPITER, this amounts to 100.3T~grid points or 501T~degrees of freedom, given its size and matching architecture to Alps.
On El~Capitan, we use $1380^3$ grid points per MI300A GPU and reach 113T~grid points using 10750~nodes (11.1K~nodes full system).
The lower grid count on Alps and El~Capitan is due purely to system size.

\begin{figure}[t]
    \centering
    \tikzsetnextfilename{strongScaling-fp32-comparison}
    \begin{tikzpicture}
    \begin{axis}[
        xmode=log,
        ymode=log,
        xtick={8,16,32,64,128,256,512,1024,2048,4096,8192,16384},
        ytick={1, 4, 16, 64, 256, 1024},
        log basis x={2},
        log basis y={2},
        ylabel={Speedup},
        xlabel={Number of Nodes},
        width=2.75in,
        height=0.65in,
        xmin=8,
        xmax=2^14,
        ymin=1,
        ymax=2^11,
        legend style={legend cell align=left,at={(0.5,1.0)},
            anchor=south,draw=none,fill=none,
            inner sep=5pt,legend columns=4,
            /tikz/every even column/.append style={column sep=0.1cm},
        },
        ]

        \begin{scope}[on background layer]
            \draw[fill=black!10,draw=none] 
                (axis cs:9450,1) rectangle (axis cs:2^14,2^11); 
            \addplot +[mark=none,very thick,black,forget plot] coordinates {(9450, 1) (9450, 2^11)};
        \end{scope}

        % ideal
        \addplot[black,dashed,very thick, domain=8:9450, samples=10] {x/8};
        \addlegendentry{Ideal}
        
        \addlegendimage{%
            area legend,%
            fill=black!10,%
            draw=none,%
            legend image code/.code={%
              \path[fill=black!10] (0cm,-0.1cm) rectangle (0.6cm,0.15cm);
              \draw[very thick,black] (0cm,-0.1cm) -- (0.0cm,0.15cm);
            }
        }
        \addlegendentry{Full System};
        
        % Frontier single
        \addplot[darkrust,thick,mark=*,forget plot] table[col sep=comma,header=false,x index=0,y index=3] {figures/strongScaling/frontierSingleUnified.csv};
        \addlegendimage{thick,mark=*,darkrust};
        \addlegendentry{This work};
        
        % \addplot[seagreen,thick,mark=*,forget plot] table[col sep=comma,header=false,x index=0,y index=3] {figures/strongScaling/frontierSingleNoUnified.csv};
        
        \addplot[thick,mark=square*,mark size=2pt,color=seagreen] coordinates {
            (8, 1.00)
            (16, 1.96)
            (32, 3.60)
            (64, 6.16)
            (128, 10.93)
            (256, 18.39)
            (512, 27.13)
            (1024, 37.28)
            (2048, 48.46)
            (4096, 61.83)
            (8192, 60.39)
        };
        \addlegendimage{thick,mark=square*,seagreen};
        \addlegendentry{Baseline};
        
        % \node at (axis cs:8,2048) [below right] {Frontier (Trento+MI250X; FP32)};
    \end{axis}
\end{tikzpicture}
    \Description{Plots comparing strong scaling performance of IGR to the current state-of-the-art.}
    \caption{
        Strong scaling following \cref{fig:strongScaling}, here for FP32 on Frontier.
        We compare the performance of the current work to the optimized baseline method as labeled.
        The current work uses and can accommodate 10.5B grid points per node, while the baseline accommodates 421M grid points, both of which are used for the 8-node speedup reference.
    }
    \label{fig:strongScalingcomparison}
\end{figure}

Strong scaling results on Alps, Frontier, and El~Capitan are shown in \cref{fig:strongScaling}.
The base case uses $8$ nodes ($32$ GPUs), resulting in $32$ ranks on El~Capitan and Alps and $64$ ranks on Frontier, arranged in a rectilinear configuration.
We use FP16/32 mixed precision for storage/computation; unified memory is enabled via UVM on Frontier and Alps and USM on El~Capitan; GPU-aware MPI is used on Alps. 
For a 32-fold increase in device count, we achieve strong scaling efficiencies of 90\%, 90\%, and 86\% on El~Capitan, Frontier, and Alps.
When scaling to the \textit{full systems}, we achieve 44\% (El~Capitan), 44\% (Frontier), and 80\% (Alps) strong scaling efficiencies.
Thus, one can execute an 8~node computation on the full system, decreasing time to solution by a factor of about~\textit{500}.

\Cref{fig:strongScalingcomparison} shows optimized baseline numerics in FP32 for the same problem, which do not strong-scale well compared to the current work.
We observe 6\% strong scaling efficiency for the baseline and 38\% for the current work when scaling from 8~nodes to the full Frontier system. 

\subsection{Energy efficiency}

\Cref{tab:igr_vs_weno} shows the measured energy consumption per grid cell per time step, comparing our IGR implementation with a prior state-of-the-art WENO implementation.
The problem size is adjusted to exhaust the total GPU/APU memory on a single node in double precision.
On the NIVIDA~GH200 we measured the in-core implementation.
On the AMD~MI250X and MI300A, we use the CCE 19.0.0 compiler with GPU-only (in-core) and unified memory builds.

The IGR method significantly reduces energy consumed compared to the current state of the art on all platforms.
The most significant reduction comes from the improved time to solution, and to lowest order, the Frontier and El~Capitan energy improvements match the grind time speedups in \cref{tab:grindTimes}.
Due to the higher power draw of the WENO scheme on Alps, we observe energy savings beyond those resulting from grind time speedup.

The largest improvement is realized on Frontier with a $5.38$-times improvement in energy consumed by the GPU.
As described in \cref{sec:measurement}, the Frontier measurements include only GPU and GPU HBM energy and do not represent changes that may have impacted energy consumption of the CPU or CPU memory.

\begin{table}[t]
    \centering
    \caption{
    Energy in \si{\micro\joule} per grid cell per time step for the baseline implementation compared to the current work.
    }
    \begin{tabular}{ r r r r }
        \toprule
       Energy (\textmu{}J)      & \makecell{El~Capitan \\ (MI300A)}    & \makecell{Frontier \\ (MI250X)}  & \makecell{Alps \\ (GH200)} \\ 
       \midrule
       Baseline & 15.24 & 10.67 & 9.349 \\
       IGR      & 3.493 & 1.982 & 2.466 \\
       \bottomrule
    \end{tabular}
    \label{tab:igr_vs_weno}
\end{table}

\section{Implications}

This work presents a highly scalable technique for predictive simulation of compressible fluid flows.
We demonstrated the method's capability by simulating high-Mach many-rocket engine thrusters and their plume--plume interaction.
This demonstrates that fully resolved simulations of these systems are within reach of current supercomputers, exceeding the scales demanded of state-of-the-art current and prototype spacecraft. 
This advance paves the way for the computation-driven design of critical components for space exploration. 

The combination of IGR, the resulting simplified algorithm, and its optimized implementation improves the efficiency of simulating compressible flow, as measured by key metrics: time to solution, memory footprint, and energy to solution. 
We leverage the closely coupled architectures of modern supercomputing platforms, including the MI250X, MI300A, and GH200, which serve as proof of concept for the efficient use of unified memory addressing. 

Furthermore, our simplified algorithm is amenable to mixed-precision computation.
The resulting computational savings enable the first CFD simulations with more than \emph{200~trillion} grid points and \emph{1~quadrillion degrees} of freedom, exceeding the previous largest such simulations by a factor of 20.
At the same time, compared to the baseline, the improved grind times and near-ideal strong scaling of our approach enable flagship supercomputers to achieve orders of magnitude reductions of time to solution on smaller problems.
This unlocks new opportunities for integrating simulation into design optimization.
Beyond flagship supercomputing, the drastic increase in grid points per device extends the capabilities of resource-constrained users. 

This work focuses on a set of thruster plumes.
However, the methodology shown is suited to general multiphase and multicomponent flows, spanning fields ranging from biomedical treatments to marine and aviation applications.
Our work thus has the potential to aid the understanding of a broad range of physical phenomena.
IGR is \textit{agnostic} to the numerical discretization, as it regularizes the momentum balance equation of the associated PDE.
Thus, one could apply the same strategy to other techniques, such as discontinuous Galerkin or finite difference methods.

This work demonstrates how users can avoid expensive and complex viscous, nonlinear numerics via IGR.
This achieves markedly better scalability, speed, problem sizes, and time- and energy-to-solution than current state-of-the-art methods.
The spatial discretizations used in this work adhere to traditional algorithmic design patterns commonly used in computational fluid dynamics.
Removing the need for numerical shock capturing enables a vast and largely unexplored design space for numerical methods for fluid dynamics problems.

\section{Acknowledgments}
 
SHB acknowledges the use of resources of the Oak Ridge Leadership Computing Facility at the Oak Ridge National Laboratory, which is supported by the Office of Science of the U.S. Department of Energy under Contract No. DE-AC05-00OR22725 and allocation CFD154 (PI~Bryngelson).
This work was also supported by a grant from the Swiss National Supercomputing Centre (CSCS) for Alps.

FS acknowledges support from the Air Force Office of Scientific Research under award number FA9550-23-1-0668 (Information Geometric Regularization for Simulation and Optimization of Supersonic Flow).

The authors gratefully acknowledge contributions from Scott~Futral (LLNL), Rob~Noska (HPE), Michael~Sandoval (OLCF), and Mat~Colgrove (NVIDIA).

\bibliographystyle{ACM-Reference-Format}
\bibliography{main}

\end{document}